\documentclass[aps,prd,reprint,twocolumn,superscriptaddress,nofootinbib,floatfix]{revtex4-2}

\usepackage{amsmath,amssymb}
\usepackage{graphicx}
\usepackage{hyperref}
\usepackage{color}
\usepackage{physics}
\usepackage{slashed}
\usepackage{booktabs}
\usepackage{comment}
\usepackage{subcaption}
\usepackage{soul}
\usepackage{array}
\usepackage[utf8]{inputenc}

\allowdisplaybreaks

\begin{document}
\preprint{UMN-TH-4526/26}

\title{Electroweak Restoration: SMEFT and HEFT}

\author{Ian M. Lewis}
\email{ian.lewis@ku.edu}
\affiliation{Department of Physics and Astronomy, University of Kansas, Lawrence, Kansas, 66045~ U.S.A.}

\author{Zhen Liu}
\email{zliuphys@umn.edu}
\affiliation{%
  School of Physics and Astronomy, University of Minnesota, Minneapolis, MN 55455, USA%
}
\author{Ishmam Mahbub}
\email{mahbu008@umn.edu}
\affiliation{%
  School of Physics and Astronomy, University of Minnesota, Minneapolis, MN 55455, USA%
}
\begin{abstract}
Colliders continue to push our understanding of electroweak (EW) interactions to ever higher energies. At high energies, many observables in the broken EW theory are expected to approach the unbroken theory. This is the electroweak restoration regime, where longitudinal gauge boson production corresponds to Goldstone boson production. As such, in this paper we investigate electroweak restoration in the context of linear and non-linear realizations of the EW symmetry in longitudinal di-boson production: $f\bar{f}'\rightarrow V_LV'_L$ and $f\bar{f}'\rightarrow V_Lh$, where $V_L,V'_L$ are longitudinal gauge bosons, $h$ is the Higgs boson, and $f,f'$ are SM fermions. For the linear beyond the SM (BSM) theory, we use the Standard Model Effective Theory (SMEFT), and for the non-linear BSM theory, we use Higgs Effective Field Theory (HEFT). We give a general discussion of these amplitudes and cross sections in the SM, SMEFT, and HEFT. 
Using the Goldstone boson equivalence theorem, we derive a set of ratios among high energy $V_LV'_L$ and $V_Lh$ amplitudes that are expected to approach one in the SM and SMEFT. Through our explicit calculations, we show that these ratios do indeed approach one in the SM and dimension-6 SMEFT, but not necessarily HEFT. 
Beyond the amplitudes, both theoretically and experimentally, we identify the cross section ratio of $W^\pm_LZ_L$ and $W^\pm_L h$ as a particularly promising observable to probe EW restoration and distinguish HEFT and SMEFT. 
Using current LHC measurements as well as projections for $W^\pm_L h$ measurements, we project HL-LHC sensitivities to probing the linear vs. non-linear realizations of the EW symmetry. 
\end{abstract}

\maketitle

\section{Introduction}

The Large Hadron Collider (LHC) continues to push our understanding of electroweak (EW) physics and EW symmetry breaking (EWSB) to ever higher energies. At these high energies ($E \gg m_W,\,m_Z,\,m_h$), EW hard scattering approaches the regime  where the weak-boson and Higgs masses enter amplitudes only through power-suppressed corrections of order $m/E$. 
In this limit, amplitudes with longitudinally polarized gauge bosons converge to the corresponding Goldstone-boson amplitudes via the Goldstone boson equivalence theorem (GBET)~\cite{Cornwall:1974km,Lee:1977eg,Chanowitz:1985hj,Gounaris:1986cr,Bagger:1989fc,PhysRevD.41.2294,He:1992nga,Cuomo:2019siu}. This is the so-called {\it electroweak restoration} regime: the amplitudes in the broken EW theory approach those in the unbroken theory. 
The leading high energy behavior of helicity amplitudes in the Standard Model (SM) are controlled by the restored $SU(2)_L\times U(1)_Y$ gauge symmetry and by the doublet structure of the scalar sector, which correlates the physical Higgs with the three would-be Goldstone bosons. Importantly, correlations among amplitudes with longitudinal gauge boson final states provide a test of the doublet nature of the Higgs field, and therefore the symmetry structure of the scalar sector. 

A probe of EW restoration at the LHC was proposed via a test of convergence in Ref.~\cite{Huang:2020iya}. There, EW restoration is defined by comparing $Vh$ production in the broken theory to the corresponding $Gh$ rate in the EW symmetric theory, where $V=W^\pm,Z$ are massive EW gauge bosons and $G=G^\pm,G^0$ are the corresponding Goldstone bosons.
One introduces the differential signal strength
\begin{equation}
\mu_{Vh}(p_T^h)
\;=\;
\frac{d\sigma(pp\to Vh)/dp_T^h}{d\sigma(pp\to Gh)/dp_T^h|_{v=0}}
,
\label{eq:muvh_def}
\end{equation}
where \(p_T^h\) is the Higgs transverse momentum,  $d\sigma(pp\rightarrow Gh)$ is calculated in the unbroken EW theory, and $d\sigma(pp\rightarrow Vh)$ is calculated in the broken EW theory. Electroweak restoration is then probed via an empirical test of convergence through  \(\mu_{Vh}(p_T^h)\to 1\) at large \(p_T^h\).
These ideas are extended to muon colliders in Ref.~\cite{MuCsEWRestoration}, where electroweak restoration is studied in the \(Zh\), \(WW\), and \(WW\gamma\) channels. The analysis tests the convergence of longitudinal gauge-boson rates to the corresponding Goldstone predictions at \(\sqrt{S}=1,\,3,\) and \(10~\mathrm{TeV}\), including simulations of electroweak-boson decays and detector-level event reconstruction. 

Ref.~\cite{Brodsky:1982sh,Baur:1994ia,Capdevilla:2024amg} proposed observation of the radiation amplitude zero (RAZ) in \(\,f\,\bar{f}'\to WZ\,\). In the high energy limit, restored gauge symmetry enforces a vanishing transverse amplitude \(W_T Z_T\) at a specific center-of-mass scattering angle, i.e. radiation amplitude zero. At the RAZ, the di-boson amplitudes are naturally dominated by the longitudinal modes. Hence, EW restoration can be probed at the RAZ.

Those previous studies~\cite{Huang:2020iya,Capdevilla:2024amg} focused on the case of the SM. The presence of beyond the SM (BSM) physics can alter the nature of EW restoration. In this work, we extend the previous studies by analyzing EW restoration in the SM effective field theory (SMEFT) \cite{Buchmuller:1985jz, Grzadkowski:2010es, Jenkins:2013zja, Falkowski:2014tna,Berthier:2016tkq, Brivio:2017vri,Brivio:2017bnu} and the Higgs effective field theory (HEFT) \cite{Feruglio:1992wf,Burgess:1999ha,Grinstein:2007iv,Buchalla:2012qq,Alonso:2012px,Buchalla:2013rka,Brivio:2013pma,Brivio:2016fzo,Alonso:2015fsp,Alonso:2016oah}. In SMEFT the EW symmetry is linearly realized and in HEFT it is generically non-linearly realized.
Hence, the nature of EW symmetry is expected to be different in the two theories. We will show that the \emph{ratios} of di-boson scattering amplitudes and cross sections with longitudinally polarized gauge bosons in the presence of SMEFT and HEFT can distinguish linear and non-linear realizations of EW symmetry.\footnote{We note that there have been studies comparing SMEFT and HEFT in multi-Higgs production. See~\cite{Liu:2018vel,Liu:2018qtb,Chang:2019vez,Liu:2019rce,Abu-Ajamieh:2020yqi,Gomez-Ambrosio:2022qsi,Domenech:2022uud,Englert:2023uug,Delgado:2023ynh,Liu:2023jbq,Remmen:2024hry,Domenech:2025gmn} and references therein.}

In the EW restoration regime, longitudinal gauge bosons in the broken EW theory are expected to correspond to Goldstone bosons in the unbroken EW theory. In the linear realization of EW symmetry, the Higgs and Goldstone bosons form a complex $SU(2)_L$ doublet, while in the non-linear realization, the Goldstone bosons form a real $SU(2)_L$ triplet and the Higgs boson a gauge singlet scalar.  Hence, the linear realization of EW symmetry is expected to predict (approximate) relationships between the Higgs and Goldstone boson amplitudes at high energies, while in the non-linear EW symmetry there are no relationships between the Goldstone and Higgs amplitudes\footnote{While in HEFT it may be expected that there are relations between the different Goldstone boson amplitudes, as we will show there are operators that break custodial symmetry and contribute quadratic energy growth to the amplitudes. 
}. 
Using the GBET, the relationships among Goldstone boson amplitudes imply similar relationships for amplitudes with longitudinal gauge bosons. Hence, for a complete study of EW restoration, all longitudinally polarized di-boson processes should be studied. As such, we first determine the amplitudes for $f\bar{f}\rightarrow W_L^+W_L^-$, $f\bar{f}'\rightarrow W_L^\pm Z_L$, $f\bar{f}'\rightarrow W_L^\pm h$, and $f\bar{f}\rightarrow Z_Lh$,\footnote{For the process $f\bar{f}\rightarrow Z_LZ_L$, the Goldstone bosons decouple in both the SM and SMEFT, in the massless fermion limit.} where the subscript $L$ indicates longitudinal polarization, in both SMEFT and HEFT and compare their high energy behavior. 

Additionally, for massless fermions, in the SM with an unbroken EW theory the pair production of Goldstone bosons via fermion-antifermion annihilation proceeds through the $s$-channel exchange of hypercharge and $SU(2)_L$ gauge bosons currents. Since the broken EW symmetry converges to the unbroken symmetry at high energies, we use the products of quark and Higgs doublet currents and the GBET to identify a set of ratios of SM amplitudes of $f\bar{f}'\rightarrow V_LV'_L$ and $f\bar{f}'\rightarrow V_Lh$ and their linear combinations that are predicted to converge to one at high energies. Using explicit calculations and the structure of Warsaw basis SMEFT operators~\cite{Grzadkowski:2010es}, we show that the quadratic energy growth in dimension-6 SMEFT is expected to obey the same relations as the SM. Since the Higgs boson is a singlet in the non-linear realization, HEFT in general does not obey all of the same set of amplitude relationships. 
Measuring these ratios would be a robust test of electroweak restoration and the structure of EW symmetry. 

We also compare the ratios of the partonic cross sections. As we show, the ratio of $W^\pm_LZ_L$ and $W^\pm_Lh$ rates is particularly sensitive to linear vs. non-linear EW symmetry due to the ratio being predicted to approach one in the SM and SMEFT even after the initial state quark helicities are summed.  In HEFT, even small non-SM contributions to the amplitudes cause this ratio of rates to diverge at high energy. 
We note that while we focus on quark-antiquark initial states, the GBET analysis in the SM and SMEFT is applicable to any fermion-antifermion initial state.

From this discussion, our main phenomenological focus in this work is the ratio of \(W^\pm_LZ_L\) to \(W^\pm_Lh\) cross sections in the LHC. This choice is also well motivated experimentally 
since ATLAS has already observed the simultaneous production of longitudinally polarized \(W^\pm\) and \(Z\) bosons with a significance of \(7.1\sigma\) in the fully leptonic channel. This corresponds to a measurement of the joint longitudinal fraction \(f_{00}=0.067\pm0.010\) \cite{ATLAS:2022oge,ATLAS:2024qbd}. For \(W^\pm h\), recent analyses in the \(\ell^\pm \nu \gamma\gamma\) final state project an approximately \(10\%\) precision measurement of the inclusive \(W^\pm_L h\) cross section at \(3000~\mathrm{fb}^{-1}\) \cite{Colyer:2025ehv}. Together with the theoretical motivation, these developments make the \(W^\pm_LZ_L/W^\pm_Lh\) cross section ratio a natural observable for probing electroweak restoration experimentally.

In Sec.~\ref{sec:theory} we give an overview of the relevant HEFT and SMEFT operators under study, as well as the amplitudes for longitudinal di-boson final states. In Sec.~\ref{sec:results} we discuss the ratios of di-boson amplitudes in SM, SMEFT, and HEFT, including predictions from the GBET for the high energy behaviour of SM and dimension-6 SMEFT amplitudes. We also discuss the di-boson partonic cross sections, showing that the ratio of $W^\pm_LZ_L$ and $W^\pm_L h$ cross sections are a good test of linear vs. non-linear realizations of EW symmetry.  In Sec.~\ref{sec:LHCresults} we project HL-LHC  sensitivity to distinguishing the linear vs. non-linear EW symmetry comparing $W^\pm_LZ_L$ to $W^\pm_Lh$. We conclude in Sec.~\ref{sec:conc}.

\section{Theory}
\label{sec:theory}

In this section, we discuss EW restoration for quark-antiquark annihilation into longitudinal di-bosons: 
\begin{alignat*}{2}
q\bar{q}&\rightarrow W^+_LW^-_L,\quad &q\bar{q}'&\rightarrow W^\pm_L Z_L\\
q\bar{q}&\rightarrow Z_Lh,\,&q\bar{q}'&\rightarrow W^\pm_L h,
\end{alignat*}
where the subscript $L$ indicates longitudinal polarization\footnote{Note that the projection into massive gauge boson helicity states depends on the choice of spin-axis and is not unique. Throughout this work, we choose the helicity basis in the partonic center of momentum frame.}.

As mentioned previously, we choose SMEFT for the BSM model with a linear realization of EW symmetry and HEFT for the BSM model with a non-linear EW symmetry. Both SMEFT and HEFT have many operators that can contribute to di-boson production. For both HEFT and SMEFT amplitudes, we work at linear order in the Wilson coefficients. We will consider all dimension-6 SMEFT operators that have a quadratic energy growth at the amplitude level. We also consider one additional bosonic SMEFT operator that breaks {\rm custodial symmetry}.

For HEFT, we will restrict ourselves to a subset of operators that produce quadratic energy growth at the amplitude level, sufficiently illustrating the differences between the two theories.  
All amplitudes are presented as $\mathcal{M}(q_\lambda \bar{q}'_{\lambda'}\rightarrow V_LV'_L/V_Lh)$ where $\lambda$ is the quark helicity and $\lambda'$ is the antiquark helicity.

\subsection{SM Amplitudes}
\label{sec:SMAmp}

While the amplitudes for di-boson production in the SM are well-known~\cite{Gaemers:1978hg,Duncan:1985vj,Hagiwara:1986vm,Baglio:2017bfe,Baglio:2018rcu,Huang:2020iya}, for completeness and comparative analysis with later EFT modifications, we report them here. In the Standard Model, the high energy limit of 
$W^+_L Z_L$, $W^+_L W^-_L$, $W^+_L h$, and $Z_L h$ are 

\begin{subequations}
\label{eq:qqWZWh_helicity}
\begin{align}
&\mathcal{M}_{\rm SM}\bigl(q_- \bar q'_+ \to W_L^\pm Z_L\bigr) 
    = \pm\,2   m_W^2   G_F \sin\theta
    + \mathcal{O}\!\left(\frac{m^2}{E^2}\right), \\[4pt]
 &\mathcal{M}_{\rm SM}\bigl(q_- \bar q_+ \to W_{L}^+ W_{L}^-\bigr)\\
 &\quad =-2\sqrt{2} {m}_Z^2 {G}_F\left( {c}_W^2 T_3^q+ {s}_W^2Y_L^q\right)\sin\theta + \mathcal{O}\!\left(\frac{m^2}{s}\right), \nonumber\\[4pt]
 &\mathcal{M}_{\rm SM}\bigl(q_+ \bar q_- \to W_{L}^+ W_{L}^-\bigr)
 =2\sqrt{2} {m}_Z^2 {G}_F {s}_W^2Y_R^q\sin\theta\nonumber\\
    & \hspace{15em}+ \mathcal{O}\!\left(\frac{m^2}{s}\right), \\[4pt]
&\mathcal{M}_{\rm SM}\bigl(q_- \bar q'_+ \to W_L^\pm h\bigr) 
  = 2   m_W^2   G_F \sin\theta
    + \mathcal{O}\!\left(\frac{m^2}{s}\right),\\[4pt]
&\mathcal{M}_{\rm SM}\bigl(q_{-} \bar q_{+} \to Z_L h\bigr) 
=2\sqrt{2} {G}_F {m}_Z^2\, {g}_L^{Zq}\sin\theta\nonumber\\
    & \hspace{15em} + \mathcal{O}\!\left(\frac{m^2}{s}\right), \\[4pt]
  &\mathcal{M}_{\rm SM}\bigl(q_{+} \bar q_{-} \to Z_L h\bigr) 
  =-2\sqrt{2} {G}_F {m}_Z^2\, {g}_R^{Zq}\sin\theta\nonumber\\
    & \hspace{15em} + \mathcal{O}\!\left(\frac{m^2}{s}\right),
\end{align}
\end{subequations}
where, for the $W^+W^-/W^+Z/W^+h$ processes, $\theta$ is the angle of $W^{+}$ with respect to the direction of the incoming quark
in the partonic center of momentum frame; for the $W^-Z/W^-h$ processes, $\theta$ is the angle between the $W^-$ and initial state quark; for $Zh$ production $\theta$ is the angle between the $Z$ and initial state quark; $\sqrt{s}$ is the
partonic center-of-mass energy; $m$ is an EW scale mass; $ {g}_L^{Zq}=T_3^q- {s}_W^2 Q$ is the left-handed quark coupling to a $Z$; $ {g}_R^{Zq}=- {s}_W^2 Q$ is the right-handed quark coupling to the $Z$; $Y_L^q=1/6$ is the left-handed quark hypercharge; the hypercharge for right-handed quarks are $Y^u_R=2/3,Y^d_R=-1/3$; $Q$ is the initial state quark charge; $T_3^u=+1/2$; $T_3^d=-1/2$; ${s}_W=\sin {\theta}_W, {c}_W=\cos {\theta}_W$ with $ {\theta}_W$ being the weak mixing angle; $m_W$ is the $W$-mass; $m_Z$ is the $Z$-mass, and $G_F$ is the Fermi decay constant. 
Here, we write the amplitude in the massless quark limit.
For longitudinal vector boson final states, all other helicity amplitudes not shown here are suppressed by at least
 $\mathcal{O}(m/\sqrt{s})$ in the leading order.

\subsection{SMEFT Operators and Amplitudes}
\label{sec:SMEFTAmp}

In SMEFT, the EW symmetry is linearly realized. That is, SMEFT operators are built out of the SM Higgs doublet $H$:
\begin{eqnarray}
H=\frac{1}{\sqrt{2}}\begin{pmatrix}-\sqrt{2}\,i\,G^+\\v+h+i\,G^0\end{pmatrix},
\label{eq:H}
\end{eqnarray}
where $G^+,G^0$ are the Goldstone bosons, $h$ is the Higgs boson, and $v=246$~GeV is the SM-like Higgs vacuum expectation value (vev). We use the Warsaw basis~\cite{Grzadkowski:2010es} and consider the following five dimension-6 SMEFT operators 
\begin{subequations}
\label{eq:dim6_ops}
\begin{align}
\mathcal{Q}_{Hq}&=
        (H^\dagger i\overleftrightarrow{D}_\mu H)\,
        (\bar q_R  \gamma^\mu q_R)\,, \label{eq:QHq}\\
\mathcal{Q}_{Hq}^{(1)} &=
        (H^\dagger i\overleftrightarrow{D}_\mu H)\,
        (\bar Q_L  \gamma^\mu Q_L)\,, \\
\mathcal{Q}_{Hq}^{(3)} &=
        (H^\dagger i\overleftrightarrow{D}^{\,I}_\mu H)\,
        (\bar Q_L \sigma^I \gamma^\mu Q_L)\,, \\
\mathcal{Q}_{Hud}&=(\widetilde{H}^\dagger i\overleftrightarrow{D}_\mu H)\,
        (\bar u_R  \gamma^\mu d_R)\, ,\label{eq:QHud}\\
\mathcal{Q}_{HD} &=
        (H^\dagger D_\mu H)^*\,(H^\dagger D_\mu H)\,,
\end{align}
\end{subequations}
where $q_R$ is either a right-handed up- or down-type quark $SU(2)_L$ singlet, $Q_L$ is the left-handed $SU(2)_L$ quark doublet, and $\sigma^I$ are Pauli matrices. The covariant derivative is defined as  
$D_\mu H=(\partial_\mu + i\,g\,\frac{\sigma^I}{2}W^I_\mu + i\,g'YB_\mu)H$, $H^\dagger \overleftrightarrow{D}_\mu H= H^\dagger D_\mu H-(D_\mu H)^\dagger H$, and $H^\dagger \overleftrightarrow{D}^I_\mu H= H^\dagger \sigma^I D_\mu H-(D_\mu H)^\dagger \sigma^I H$. Finally, we perform the SMEFT computations in the $m_W,\,m_Z,\,G_F$ input parameter scheme~\cite{Brivio:2017bnu}.

Again, while these amplitudes have been known for a long time~\cite{Gaemers:1978hg,Duncan:1985vj,Hagiwara:1986vm,Zhang:2016zsp,Azatov:2016sqh,Franceschini:2017xkh,Baglio:2017bfe,Alioli:2018ljm,Baglio:2018rcu,Baglio:2018bkm,Baglio:2019uty,Baglio:2020oqu,Corbett:2023yhk}, for completeness and the sake of our discussion, we report them here. Expanding to linear order in Wilson coefficients, the leading high energy terms for the di-boson amplitudes are 
\begin{widetext}
\begin{subequations}
\label{eq:WZ-Wh-SMEFT}
\begin{align}
&\mathcal M_{\rm SMEFT}\!\bigl(q_- \bar{q}'_+ \to W_L^\pm Z_L\bigr)
= \pm\,\sqrt{2}\,C_{Hq}^{(3)}\,s\,\sin\theta
\pm \frac{  m_W^{2}}{2\sqrt{2}}\,
       \Bigl[
         C_{HD}
         + 4\,C_{Hq}^{(3)}\!\left(2 + \frac{  m_Z^{2}}{  m_W^{2}}\right)
         + 4\sqrt{2}\,  G_F
       \Bigr]\,\sin\theta  + \mathcal{O}\!\left(\frac{m^2}{s}\right),
\label{eq:WZ-LL-SMEFT}
\\[4pt]
&\mathcal M_{\rm SMEFT}\!\bigl(q_+ \bar{q}'_-\to W_L^{+} Z_L\bigr)
= \frac{C_{Hud}^*}{\sqrt{2}} \, s \,\sin\theta 
+\frac{  m_Z^2}{\sqrt{2}}\, C_{Hud}^* \, \sin\theta  + \mathcal{O}\!\left(\frac{m^2}{s}\right),
\\[4pt]
&\mathcal M_{\rm SMEFT}\!\bigl(q_+ \bar{q}'_-\to W_L^- Z_L\bigr)
= -\frac{C_{Hud}}{\sqrt{2}} \, s \sin\theta 
-\frac{  m_Z^2}{\sqrt{2}}\, C_{Hud} \, \sin\theta  + \mathcal{O}\!\left(\frac{m^2}{s}\right),
\\[4pt]
 &\mathcal{M}_{\rm SMEFT}\bigl(q_- \bar q_+ \to W_{L}^+ W_{L}^-\bigr)
    = -\,\Bigl(C_{Hq}^{(1)}+2\,T_3^qC_{Hq}^{(3)}\Bigr) s\, \sin\theta 
-  
\left[
-Q\,C_{HD}\,  m_W^2
+C_{Hq}^{(1)}\,  m_Z^2
+2\,T_3^qC_{Hq}^{(3)}\,(4  m_W^2-  m_Z^2)
\right]\sin\theta \nonumber \\&  \hspace{28em} - 2\sqrt{2} {m}_Z^2 {G}_F\left( {c}_W^2T_3^q+ {s}_W^2Y_L^q\right)\sin\theta
 + \mathcal{O}\!\left(\frac{m^2}{s}\right), \\[4pt]
 &\mathcal{M}_{\rm SMEFT}\bigl(q_+ \bar q_- \to W_{L}^+ W_{L}^-\bigr)=
    C_{Hq}\,s\, \sin\theta  +
\left[
-Q\,C_{HD}  m_W^2
+C_{Hq}  m_Z^2
\right] \sin\theta +2\,\sqrt{2} {m}_Z^2 {G}_F {s}_W^2Y^q_R\sin\theta
+\mathcal{O}\!\left(\frac{m^2}{s}\right), \\[4pt]
&\mathcal M_{\rm SMEFT}\!\bigl(q_- \bar{q}'_+\!\to\! W_L^\pm h\bigr)
= \sqrt{2}\,C_{Hq}^{(3)}\,s\,\sin\theta
+ \frac{  m_W^{2}}{2\sqrt{2}}\,
   \Biggl[-
      C_{HD}
      + 4\sqrt{2}\,  G_F
      + 4\,C_{Hq}^{(3)}\,
        \frac{2  m_W^2-  m_h^{2}}{  m_W^{2}}
   \Biggr]\,\sin\theta
   + \mathcal{O}\!\left(\frac{m^2}{s}\right),
\label{eq:Wh-LL-SMEFT}\\[4pt]
&\mathcal M_{\rm SMEFT}\!\bigl(q_+ \bar{q}'_-\!\to\! W_L^+ h\bigr)
=-\frac{C_{Hud}^*}{\sqrt{2}} s\,\sin\theta  -\frac{(2  m_W^2-  m_h^2)}{\sqrt{2}}\, C_{Hud}^* \, \sin\theta + \mathcal{O}\!\left(\frac{m^2}{s}\right)\\[4pt]
&\mathcal M_{\rm SMEFT}\!\bigl(q_+ \bar{q}'_-\!\to\! W_L^- h\bigr)
=-\frac{C_{Hud}}{\sqrt{2}} s\,\sin\theta  -\frac{(2  m_W^2-  m_h^2)}{\sqrt{2}}\, C_{Hud} \, \sin\theta + \mathcal{O}\!\left(\frac{m^2}{s}\right)\\[4pt]
&\mathcal{M}_{\rm SMEFT}\bigl(q_{-} \bar q_{+} \to Z_L h\bigr) 
=-\bigl(C_{Hq}^{(1)}-2\,T_3^qC_{Hq}^{(3)}\bigr)\,s\,\sin\theta 
+
\Bigl[
Q \,C_{HD}  m_W^2
-\bigl(C_{Hq}^{(1)}-2\,T_3^qC_{Hq}^{(3)}\bigr)\bigl(2 {m}_Z^2- {m}_h^2\bigr)
\Bigr]\,\sin\theta \nonumber \\
&\hspace{28em}+2\sqrt{2}\, {m}_Z^2\, {G}_F\, {g}_L^{Zq}\sin\theta
+\mathcal{O}\!\left(\frac{m^2}{s}\right),\label{eq:SMEFTpmZh}\\[5pt]
&\mathcal{M}_{\rm SMEFT}\bigl(q_{+} \bar q_{-} \to Z_L h\bigr) 
=C_{Hq}\,s\,\sin\theta -
\Bigl[
Q\,C_{HD}  m_W^2
-C_{Hq}\bigl(2\, {m}_Z^2-  m_h^2\bigr)
\Bigr]\sin\theta\nonumber \\
&\hspace{28em}-2\sqrt{2} {m}_Z^2 {G}_F {g}_R^{Zq}\,\sin\theta
+\mathcal{O}\!\left(\frac{m^2}{s}\right).\label{eq:SMEFTmpZh}
\end{align}
\end{subequations}
\end{widetext}
where the parameters satisfy the SM tree-level relations among couplings and masses. Note that in SMEFT the quadratic energy-growing terms in the
\(V_LV_L\) and \(V_Lh\) amplitudes are correlated, as we will discuss in
detail in the next section.

All operators $\mathcal{Q}_{Hq}$, $\mathcal{Q}_{Hq}^{(1)}$, $\mathcal{Q}_{Hq}^{(3)}$ and $\mathcal{Q}_{Hud}$ contribute to quadratic energy growth in the amplitudes, and, in the Warsaw basis, these are the only dimension-6 SMEFT operators that contribute to quadratic energy at linear order in Wilson coefficients.  $\mathcal Q_{HD}$ is a purely bosonic operator that involves only covariant derivatives of the Higgs doublet, and after electroweak symmetry breaking, it modifies the neutral electroweak sector already at tree level. In particular, it shifts the relation between the $W$ and $Z$ masses and is closely associated with the custodial-breaking $T$-parameter direction. Its effects appear in the subleading $\mathcal{O}(s^0)$ terms. There are other bosonic operators, such as $H^\dagger \sigma^IH W^I_{\mu\nu}B^{\mu\nu}$, that contribute similarly to the amplitudes. We include $\mathcal{Q}_{HD}$ for illustration purposes~\cite{deBlas:2025xhe}, in particular, comparing with a related operator in HEFT. Additionally, there are dipole and magnetic moment operators that could contribute to these processes. However, for massless initial state quarks, they would not interfere with the SM processes at linear order due to their chiral structure.  Hence, they would only contribute to quadratic order at the cross section level which is formally a dimension-8 contribution to the cross section.

\subsection{HEFT Operators and Amplitudes}
\label{sec:HEFTAmp}

For HEFT, we focus on a subset of operators that contribute to quadratic energy growth at the amplitude level. Unlike SMEFT, in HEFT, the EW symmetry is non-linearly realized. That is, the Higgs boson is treated as a gauge singlet scalar, and the Goldstone bosons are collected into a dimensionless unitary matrix transforming under global symmetry $SU(2)_L \times SU(2)_R$,

\begin{align}
   U(x) = \exp\!\left( \frac{i\,\sigma^I \pi^I(x)}{v} \right), \quad U(x)\to L\,U(x)\,R^\dagger, \label{eq:Goldstone}
\end{align}
where $v=246$~GeV is the EW vev, $\pi^I(x)$ are the three Goldstone fields \footnote{With this definition, we can identify $\sigma^\pm =(\sigma^1\pm i\sigma^2)/\sqrt{2}$ and connect $\pi^I$ with the Goldstone bosons in the Higgs doublet: $G^\pm = -(\pi^1\mp i\,\pi^2)/\sqrt{2}$, and $G^0=-\pi^3$.}, and $L$ and $R$ denote the global $SU(2)_L$ and $SU(2)_R$ transformations. In the electroweak theory, the $SU(2)_L$ is gauged and the subgroup of $SU(2)_R$ generated by $\sigma^3$ is gauged, which is identified with the hypercharge $U(1)_Y$. We focus on the HEFT operators~\cite{Brivio:2016fzo}:
\begin{subequations}
\label{eq:HEFT-ops}
\begin{align}
&\mathcal{N}_1^{Q}(h) =
  i\,\bar Q_L \gamma^\mu V_\mu Q_L\, \mathcal{F}_1(h),\label{eq:HEFTn1Q}
\\[4pt]
&\mathcal{N}_2^{Q}(h) =
  i\,\bar Q_R \gamma^\mu U^\dagger V_\mu U\, Q_R\, \mathcal{F}_2(h),
\\[4pt]
&\mathcal{N}_4^Q(h)=\bar{Q}_R \gamma_\mu U^\dagger[V^\mu,T]U Q_R F_4(h)\\[4pt]
&\mathcal{N}^Q_5(h)=i\,\bar{Q}_L\gamma_\mu \{V^\mu,T\}Q_L\,F_5(h)\\[4pt]
&\mathcal{N}_6^Q(h)=i\,\bar{Q}_R\gamma_\mu U^\dagger \{V^\mu,T\} U Q_R F_6(h)\\[4pt]
&\mathcal{N}_7^{Q}(h) =
  i\,\bar Q_L \gamma^\mu T V_\mu T\, Q_L\,\mathcal{F}_7(h), \\[4pt]
  &\mathcal{N}_8^{Q}(h) =
  i\,\bar Q_R \gamma^\mu U^\dagger T V_\mu T U\, Q_R\, \mathcal{F}_8(h) \\[4pt]
  &\mathcal P_3(h) = \frac{i}{4\pi}\Tr \, \!\bigl(W_{\mu\nu}[V^\mu,V^\nu]\bigr)\, \mathcal{F}_3(h).\label{eq:HEFTp3}
\end{align}
\end{subequations}
following SM notation to denote the $SU(2)_L \times U(1)_Y$ gauge fields, where $V_\mu = (D_\mu U)U^\dagger$, $T = U\sigma^3 U^\dagger$, $D_\mu U=\partial_\mu U+i\,\frac{g}{2}\,\sigma^I W_\mu^I\,U-i\,\frac{g'}{2}\,U\,\sigma^3\,B_\mu$, and $W_{\mu\nu}=W_{\mu\nu}^J \sigma^J/2$ is the $SU(2)_L$ field strength tensor.

The operators in Eq. \ref{eq:HEFT-ops} are built from HEFT covariant building blocks and are
invariant under the nonlinearly realized electroweak gauge symmetry. \(\mathcal N_1^Q\), \(\mathcal N_5^Q\), and \(\mathcal N_7^Q\) modify the interactions of \(Q_L\) with the electroweak Goldstone and gauge sector, while \(\mathcal N_2^Q\), \(\mathcal N_4^Q\), \(\mathcal N_6^Q\), and \(\mathcal N_8^Q\) involve right-handed quark interactions.  \(\mathcal P_3(h)\) is a purely bosonic operator, and modifies the gauge and Goldstone self-interactions.

Note that the insertions of \(T=U\sigma^3U^\dagger\) differentiate neutral and charged electroweak components. For example, in the leading order in the Goldstone expansion, \(T\simeq \sigma^3\), so \(T\sigma^3T\simeq \sigma^3\) while \(T\sigma^\pm T\simeq -\sigma^\pm\). Furthermore, among the operators considered, \(\mathcal N_4^Q\) is CP-odd~\cite{Brivio:2016fzo}.

As mentioned previously, in HEFT, the physical Higgs boson is treated independently from the Goldstone matrix, and so the operators are multiplied by generic functions of the singlet field $\mathcal{F}_i(h)$:
\begin{equation}
\mathcal{F}_i(h)=\left(1+2\kappa_i\,\frac{h}{v}+\kappa_i^{(2)}\,\frac{h^2}{v^2}+\mathcal O(h^3)\right) .
\end{equation}

The terms proportional to $\kappa_i$ will produce a four-point contact term relevant for $q \bar{q}' \rightarrow V h$ and grow as $E^2$ when the vector boson is longitudinally polarized. Due to $\kappa_i$ being an independent parameter, these explicitly break the correlation between \(q\bar q'\to V_L V'_L\) and \(q\bar q'\to V_L h\) processes.

High-energy expansions of amplitudes with longitudinal di-boson final states are 
\begin{widetext}
\begin{subequations}
\begin{align}
&\mathcal M_{\rm HEFT} \bigl(q_- \bar{q}'_+ \!\to\! W_L^{\pm} Z_L\bigr)
=
\pm \sin\theta \Bigg[
\frac{  G_F}{\pi}\,s
\left(
4\pi(n_1^Q+n_7^Q)-\frac{  g}{2}\,c_3
\right)
+\frac{  G_F}{\pi}\,\frac{  g}{2}\,c_3\,(  m_Z^2- 2   m_W^2)
\nonumber\\[2pt]
&\hspace{16em}
+4  G_F  m_Z^2 n_7^Q
+4  G_F(2  m_W^2+  m_Z^2)n_1^Q
+2  m_W^2  G_F
\Bigg]
+\mathcal O\!\left(\frac{m^2}{s}\right), \label{eq:HEFTWZ}\\[4pt]
&\mathcal M_{\rm HEFT} \bigl(q_+ \bar{q}'_- \!\to\! W_L^{\pm} Z_L\bigr)
=  \pm \Bigl(4  G_F(n_2^Q-n_8^Q \mp \,2i\,n_4^Q)\, s \, \sin\theta +4\,  G_F\,  m_Z^2\,(n_2^Q-n_8^Q \mp  2\,i\,n_4^Q)\,\sin\theta \Bigr)+\mathcal{O}\!\left(\frac{m^2}{s}\right),
\\[4pt] 
&\mathcal{M}_{\rm HEFT}\bigl(q_- \bar q_+ \to W_L^+ W_L^-\bigr)
=\sqrt{2}\,  G_F\,s \,\sin\theta
\left[ 4\,n_5^Q+T_3^q\left(\frac{  g}{2 \pi}
\,c_3\
-4\,(n_1^Q-3n_7^Q)
\right)\right]\\[4pt]
&+\sqrt{2}\,  G_F\,\sin\theta \left[
\frac{  g}{2 \pi}\,
\Bigl((3Q-2T_3^q)  m_W^2+3  m_Z^2(T_3^q-Q)\Bigr)c_3
+\,4 T_3^q\Bigl[(  m_Z^2-4  m_W^2)\,n_1^Q+(4  m_W^2+  m_Z^2)\,n_7^Q\Bigr]+4\, {m}_Z^2\,n_5^Q
\right]\nonumber\\[4pt]
&\hspace{24em}-2\sqrt{2} {m}_Z^2 {G}_F\left( {c}_W^2T_3^q+ {s}_W^2Y_L^q\right)\sin\theta+ \mathcal O\!\left(\frac{m^2}{s}\right),\nonumber \\[4pt] 
&\mathcal{M}_{\rm HEFT}\bigl(q_+ \bar q_- \to W_L^+ W_L^-\bigr)=-
\sqrt{2}  G_F\sin\theta
\Bigg(
4 \left(n_6^Q+ T_3^q  \bigl(n_2^Q+n_8^Q\bigr)\right) \bigl(s+  m_Z^2\bigr)
+Q \bigl(  m_W^2-  m_Z^2\bigr)
 \left[
2
+ \frac{3   g}{2\pi}c_3
\right]
\Bigg)
+\mathcal{O}\!\left(\frac{m^2}{s}\right),\\[4pt]
 &\mathcal M_{\rm HEFT}\bigl(q_- \bar q'_+ \!\to\! W_L^{\pm} h\bigr)
= 4\,  G_F\,\sin\theta\,(\kappa_1 n_1^Q-\kappa_7 n_7^Q)
 \Bigl[
s-  m_h^2+  m_W^2
\Bigr]
+2\,  G_F\,  m_W^2\,\sin\theta\,(2n_1^Q-2n_7^Q+1) 
+\mathcal{O}\!\left(\frac{m^2}{s}\right),
\label{eq:HEFTWh} \\[4pt] 
 &\mathcal M_{\rm HEFT}\bigl(q_+ \bar{q}'_- \!\to\! W_L^{\pm} h\bigr)
= -4 \,  G_F\,\sin\theta \,(\kappa_2 n_2^Q- \kappa_8 n_8^Q\mp 2\,i\,\kappa_4n_4^Q) \Bigl[  s-   m_h^2 +  m_W^2  \Bigr]  \nonumber\\
&\hspace{24em}-4\, {G}_F {m}_W^2 \sin \theta \left(n_2^Q-n_8^Q \mp 2\,i\,\,n_4^Q\right) + \mathcal{O}\!\left(\frac{m^2}{s}\right),\\[4pt] 
&\mathcal{M}_{\rm HEFT}\bigl(q_{-}\bar q_{+}\to Z_L h\bigr)=
\sqrt{2}\,  G_F\,\sin\theta
\bigg[4\left(\kappa_5 n_5^Q+ T_3^q
\bigl(\kappa_1 n_1^Q+\kappa_7 n_7^Q\bigr)\right)
  \Bigl[s-   m_h^2 +   m_Z^2\Bigr]
+4   m_Z^2\left(n_5^Q+T_3^q\bigl(n_1^Q+n_7^Q\bigr)\right)
\bigg] \nonumber \\
& \hspace{24em}+ 2\sqrt{2} {G}_F {m}_Z^2\, {g}_L^{Zq}\sin\theta
+\mathcal O\!\left(\frac{m^2}{s}\right),
\label{eq:HEFTmpZh} \\[4pt] 
&\mathcal{M}_{\rm HEFT}\bigl(q_{+} \bar q_{-} \to Z_L h\bigr) 
= 
-4\,\sqrt{2}\,  G_F\,\sin\theta
\bigg[\left(\kappa_6 n_6^Q+T_3^q
\bigl(\kappa_2 n_2^Q+ \kappa_8 n_8^Q\bigr)\right)
 \,\Bigl[s-  m_h^2+  m_Z^2\Bigr]
+  m_Z^2\left(n_6^Q+T_3^q\bigl(n_2^Q+n_8^Q\bigr)\right)\bigg] \nonumber\\
 &\hspace{24em} -\, 2\sqrt{2} {G}_F {m}_Z^2\, {g}_R^{Zq}\sin\theta
+\mathcal O\!\left(\frac{m^2}{s}\right).
\label{eq:HEFTpmZh}
\end{align}
\end{subequations}
\end{widetext}
where $  g/2 = 2^{1/4}\,  {m}_W \,\sqrt{  {G}_F}$. The key difference between the HEFT and SMEFT amplitudes is that the leading coefficients of the \(s\sin\theta\) terms in
\(V_LV_L\) and \(V_Lh\) channels are not correlated. In generic HEFT,  there can also be additional one-loop corrections of order $s \sin \theta$ to the above amplitudes. Since we illustrate the loss of correlation among $V_L V_L$, $V_Lh$ in the energy regime where $m^2/s$ is small, we neglect the explicit loop logarithms and interpret the coefficients as renormalized effective parameters. \footnote{For the expansion in Eq. \ref{eq:Goldstone}, a \textit{generic} nonlinear Goldstone sector without additional fine-tuning is expected to become strongly coupled at a scale of order $4 \pi v$, while $m^2/s$ corrections to the leading amplitudes become small before the cut-off.}

As mentioned previously, we see that all the HEFT operators considered contribute to quadratic energy growth in the amplitudes. The $\mathcal{O}(s)$ growth from the \(\mathcal N_i^Q\) operators can be understood from their structure.
Expanding the Goldstone matrix \(U=\exp(i\sigma^I\pi^I/v)\), one finds
\begin{equation}
V_\mu
=
\frac{i}{v}\,\partial_\mu \pi^I \sigma^I
-\frac{i}{v^2}\,\epsilon^{IJK}\,\pi^I \partial_\mu \pi^J \sigma^K
+\cdots .
\end{equation}
Therefore, the \(\mathcal N_i^Q\) operators generate four-point interactions of the form
\begin{equation}
\bar q\,\gamma^\mu q\,\pi\,\partial_\mu \pi,
\qquad
\bar q\,\gamma^\mu q\,h \, \partial_\mu \pi ,
\end{equation}
where the second type of terms comes from the linear expansion in \(\mathcal F_i(h)\). At high energy, the derivative acting on the Goldstone field contributes one power of energy, $\sqrt{s}$, and the fermion current scales as $\sqrt{s}$.
Using the equivalence theorem, this implies that \(\mathcal N_i^Q\) operators generate quadratic energy growth in amplitudes with longitudinal gauge bosons.

The $\mathcal{O}(s)$ growth from the HEFT operator \(\mathcal P_3\) can be understood by noting that
\begin{align}
[V^\mu,V^\nu]
&=
\left[\frac{i}{v}\,\partial^\mu\pi^I\sigma^I,\,
      \frac{i}{v}\,\partial^\nu\pi^J\sigma^J\right]
+\cdots
\nonumber\\
&=
-\frac{2i}{v^2}\,\epsilon^{IJK}\,
\partial^\mu\pi^I\,\partial^\nu\pi^J\,\sigma^K
+\cdots .
\end{align}
Using \(W_{\mu\nu}=W_{\mu\nu}^K \sigma^K/2\) and
\( {\rm Tr}(\sigma^I\sigma^J)=2\delta^{IJ}\), one finds
\begin{equation}
\mathcal P_3(h)
\supset
\frac{1}{2\pi v^2}\,
\epsilon^{IJK}\,
W_{\mu\nu}^K\,
\partial^\mu\pi^I\,\partial^\nu\pi^J\,
\mathcal F_3(h).
\label{eq:P3Goldstone}
\end{equation}
Thus, \(\mathcal P_3\) generates a derivative interaction of one gauge field with two Goldstones. This structure is relevant for 
\(q_-\bar q'_+ \to \pi^+\pi^0\) and \(q_-\bar q_+\to \pi^+\pi^-\). For the $s$-channel exchange, the
\(\mathcal P_3\) vertex scales as
$W_{\mu\nu}\partial\pi\partial\pi\sim s^{3/2}$ , the fermion current scales as $\sqrt{s}$, and the propagator contributes
$1/ s$. Therefore,
\begin{equation}
\mathcal M_{\mathcal P_3}\sim \sqrt{s}\times s^{3/2}\times s^{-1}\sim s.
\end{equation}
This explains the quadratic energy growth proportional to \(c_3\) in the
\(W_LZ_L\) and \(W_LW_L\) amplitudes. Note that \(\mathcal P_3\) does not generate an $\mathcal{O}(s)$ growing term at linear
order for the \(q_+ \bar q'_-\) helicity configuration. Since the $W$ boson does not couple to right-handed quarks in the SM, the $\mathcal{O}(s)$ contribution due to $c_3$ enters only at quadratic order in Wilson coefficients in these cases.

We also see why \(\mathcal P_3\) does not generate a quadratic energy contribution to \(q\bar q'\to W_L h\) and \(q\bar q'\to Z_L h\). Since the commutator \([V^\mu,V^\nu]\) starts at order \(\pi^2\), the operator \(\mathcal P_3\) always contains at least two Goldstone fields. After expanding
\(\mathcal F_3(h)=1+\kappa_3 h/v+\cdots\), one obtains interactions of the form
\(h\,W_{\mu\nu} \partial^{\mu
}\pi \partial^{\nu}\pi\), rather than \(\,W_{\mu\nu}\partial^{\mu}h  \partial^{\nu}\pi\). Hence
\(\mathcal P_3\) contributes at leading order to amplitudes with two longitudinal gauge bosons, but not to \(V_L h\) amplitudes.

\section{Discussion: Linear vs. Non-Linear Realizations and the Goldstone Boson Equivalence Theorem}
\label{sec:results}
We now discuss the ratios of amplitudes and cross sections for the di-boson processes in the various theories.

\subsection{Amplitudes}

\begin{table*}
\begin{center}
\begin{tabular}{|c||c|c|c|}\hline\hline
To $\mathcal{O}(m^2/s)$& SM & SMEFT & HEFT \\\hline\hline
$\displaystyle\frac{\mathcal{M}(q_{-}\bar{q}'_+\rightarrow W^\pm_LZ_L)}{\mathcal{M}(q_{-}\bar{q}'_+\rightarrow W^\pm_Lh)}$ & $\displaystyle \pm1$ & $\displaystyle \pm1$ &$\displaystyle
\mp\,\frac{c_3   g-8\pi\,(n_1^Q+n_7^Q)}
{8\pi\,(\kappa_1 n_1^Q-\kappa_7 n_7^Q)}$ \rule[-3ex]{0pt}{7.25ex}\\\hline

$\displaystyle\frac{\mathcal{M}(q_{-}\bar{q}_+\rightarrow W^+_LW^-_L)}{\mathcal{M}(q_{-}\bar{q}'_+\rightarrow W^\pm_Lh)}$& $\displaystyle -\sqrt{2}\left(T_3^q+ \frac{s_W^2}{c_W^2}Y_L^q\right)$
& $\displaystyle -\,\frac{C_{Hq}^{(1)}+2\,T_3^qC_{Hq}^{(3)}}{\sqrt{2}\,C_{Hq}^{(3)}}
$& $\displaystyle
\sqrt{2}\frac{8\pi\,n_5^Q+\,T_3^q\Bigl(c_3   g-8\pi\,(n_1^Q-3n_7^Q)\Bigr)}
{8\pi\,(\kappa_1 n_1^Q-\kappa_7 n_7^Q)}
$\rule[-3ex]{0pt}{7.25ex}\\\hline

$\displaystyle\frac{\mathcal{M}(q_-\bar{q}_+\rightarrow Z_L h)}{\mathcal{M}(q_{-}\bar{q}'_+\rightarrow W^\pm_L h)}$ & $\displaystyle \sqrt{2}\frac{ {g}^{Zq}_L}{ {c}_W^2} $ &  $\displaystyle -\,\frac{C_{Hq}^{(1)}-2\,T_3^qC_{Hq}^{(3)}}{\sqrt{2}\,C_{Hq}^{(3)}}
$& $\displaystyle
\sqrt{2}
\frac{\kappa_5n_5^Q+T_3^q(\kappa_1 n_1^Q+\kappa_7 n_7^Q)}
{\kappa_1 n_1^Q-\kappa_7 n_7^Q}
$\rule[-3ex]{0pt}{7.25ex}\\\hline 

$\displaystyle \frac{\mathcal{M}(q_+ \bar{q}_-\rightarrow W^+_LW^-_L)}{\mathcal{M}(q_-\bar{q}_+ \rightarrow W^+_LW^-_L)}$ & $\displaystyle -\,\frac{ {s}_W^2 Y_R^q}{ {c}_W^2T_3^q+ {s}_W^2 Y_L^q}$ &  $\displaystyle  -\,\frac{C_{Hq}}{C_{Hq}^{(1)}+2\,T_3^qC_{Hq}^{(3)}}$& $\displaystyle
-\frac{8\pi\,(n_6^Q+T_3^q(n_2^Q+n_8^Q))}
{8\pi\,n_5^Q+T_3^q(c_3   g -8\pi\,(n_1^Q-3n_7^Q))}
$\rule[-3ex]{0pt}{7.25ex}\\\hline 

$\displaystyle \frac{\mathcal{M}(q_+\bar{q}_-\rightarrow Z_Lh)}{\mathcal{M}(q_-\bar{q}_+\rightarrow Z_Lh)}$ & $\displaystyle -\,\frac{ {g}_R^{Zq}}{ {g}_L^{Zq}}$& $\displaystyle  -\,\frac{C_{Hq}}{C_{Hq}^{(1)}-2\,T_3^q\,C_{Hq}^{(3)}}$ &  $\displaystyle
-\frac{\kappa_6n_6^Q+T_3^q(\kappa_2 n_2^Q+\kappa_8 n_8^Q)}
{\kappa_5 n_5^Q+T_3^q(\kappa_1 n_1^Q+\kappa_7 n_7^Q)}
$\rule[-3ex]{0pt}{7.25ex}\\\hline 

$\displaystyle \frac{\mathcal{M}(q_+\bar{q}'_-\rightarrow W^{-(+)}_L Z_L)}{\mathcal{M}(q_-\bar{q}'_+\rightarrow W^{-(+)}_L Z_L)}$ & $0^*$ & 
$\displaystyle \frac{C_{Hud}^{(*)}}{2 C_{Hq}^{(3)}}$
& $\displaystyle
-8\,\pi\frac{n_2^Q-n_8^Q+(-)2\,i\,n_4^Q}
{c_3   g-8\pi(n_1^Q+n_7^Q)}
$\rule[-3ex]{0pt}{7.25ex}\\\hline

$\displaystyle \frac{\mathcal{M}(q_+\bar{q}'_-\rightarrow W^{-(+)}_L h)}{\mathcal{M}(q_-\bar{q}'_+\rightarrow W^{-(+)}_Lh)}$ & $0^*$ & $\displaystyle -\, \frac{C_{Hud}^{(*)}}{2\,C_{Hq}^{(3)}}$ &
$\displaystyle
-\frac{\kappa_2 n_2^Q-\kappa_8 n_8^Q+(-)2\,i\,\kappa_4 n_4^Q}
{\kappa_1 n_1^Q-\kappa_7 n_7^Q}
$\rule[-3ex]{0pt}{7.25ex}\\\hline\hline
\end{tabular}
\end{center}
\caption{\label{tab:AmpRat} Ratios of amplitudes in SM, SMEFT, and HEFT in the high energy limit $E\gg   m_W,  m_Z,  m_h$ up to $\mathcal{O}(m^2/s)$. $0^*$ notes that these quantities are zero in the massless quark limit. 
}
\end{table*}
\begin{table*}
\setlength{\tabcolsep}{1pt}   
\begin{center}
\begin{tabular}{|c||c|c|c|}\hline\hline
To $\mathcal{O}(m^2/s)$& SM & SMEFT & HEFT \\\hline\hline
$\displaystyle\frac{\mathcal{M}(q_{-}\bar{q}'_+\rightarrow W^\pm_LZ_L)}{\mathcal{M}(q_{-}\bar{q}'_+\rightarrow W^\pm_Lh)}$ & $\quad\quad\displaystyle \pm1\quad\quad$ & $\quad\quad\displaystyle \pm1\quad\quad$ &$\displaystyle
\mp\,\frac{c_3   g-8\pi\,(n_1^Q+n_7^Q)}
{8\pi\,(\kappa_1 n_1^Q-\kappa_7 n_7^Q)}$  \rule[-3ex]{0pt}{7.25ex}\\\hline

$\displaystyle\frac{\mathcal{M}(q_-\bar{q}_+\rightarrow Z_L h)-\mathcal{M}(q_{-}\bar{q}_+\rightarrow W^+_LW^-_L)}{\sqrt{2}\,\mathcal{M}(q_{-}\bar{q}'_+\rightarrow W^\pm_Lh)}$& $\displaystyle 2\,T_3^q$ & $\displaystyle 2\,T_3^q$ & $\displaystyle 
-\frac{8\pi(1-\kappa_5)n_5^Q+T_3^q\left(c_3   g
-8 \pi\left[(1+\kappa_1)n_1^Q+(\kappa_7-3)n_7^Q\right]\right)}
{ 8 \pi\,(\kappa_1 n_1^Q-\kappa_7 n_7^Q)}
$\rule[-3ex]{0pt}{7.25ex}\\\hline 

$\displaystyle\frac{\mathcal{M}(q_+\bar{q}_-\rightarrow W^+_L W^-_L)}{\mathcal{M}(q_{+}\bar{q}_-\rightarrow Z_L h)}$ & $\quad \displaystyle 1\quad$   &$\displaystyle 1$  & $\displaystyle 
\,\frac{n_6^Q+T_3^q(n_2^Q+n_8^Q)}{\kappa_6 n_6^Q+T_3^q(\kappa_2 n_2^Q+\kappa_8 n_8^Q)}$\rule[-3ex]{0pt}{7.25ex}\\\hline 

$\displaystyle \frac{\mathcal{M}(q_+ \bar{q}'_-\rightarrow W^\pm_LZ_L)}{\mathcal{M}(q_+\bar{q}'_- \rightarrow W^\pm_L h)}$ & $-$ & $\displaystyle \mp1$ & $\displaystyle 
\mp\,\frac{n_2^Q-n_8^Q\mp2\,i\,n_4^Q}{\kappa_2 n_2^Q-\kappa_8 n_8^Q\mp2\,i\,\kappa_4n_4^Q}$\rule[-3ex]{0pt}{7.25ex}\\\hline 
$\displaystyle\frac{\mathcal{M}(q_-\bar{q}_+\rightarrow Z_L h)-\mathcal{M}(q_{-}\bar{q}_+\rightarrow W^+_LW^-_L)}{\mathcal{M}(q_-\bar{q}_+\rightarrow Z_L h)+\mathcal{M}(q_{-}\bar{q}_+\rightarrow W^+_LW^-_L)}$& $-\displaystyle \frac{g^2T_3^q}{{g'}^2Y_L^q}$ & $\displaystyle -2\,T_3^q\frac{C_{Hq}^{(3)}}{C_{Hq}^{(1)}}$& $\displaystyle 
-\frac{8\pi(1-\kappa_5)n_5^Q+T_3^q\left(c_3   g
-8 \pi\left[(1+\kappa_1)n_1^Q+(\kappa_7-3)n_7^Q\right]\right)}
{ 8\pi(1+\kappa_5)n_5^Q+T_3^q\left(c_3   g
-8 \pi\left[(1-\kappa_1)n_1^Q-(\kappa_7+3)n_7^Q\right]\right)}
$\rule[-3ex]{0pt}{7.25ex}\\\hline 
\hline

\end{tabular}
\end{center}
\caption{\label{tab:AmpRatEWR} Selected ratios of amplitudes in the SM, SMEFT, and HEFT in the high energy limit, shown up to $\mathcal{O}(m^2/s)$. The dash $-$ indicates that in the massless quark limit the numerator and denominator are both zero.
}
\end{table*}

In Table~\ref{tab:AmpRat} we give the ratios of the amplitudes for SM, SMEFT, and HEFT as reported in the previous section. We observe that for both the SM and SMEFT, the ratios of $W^\pm_LZ_L$ and $W^\pm_Lh$ approach one, while this is not true in HEFT. 
In the SM the story is clear: in the high energy limit $Z_L$ can be replaced by its corresponding Goldstone boson $G^0$.  Hence, the matrix elements for $W^\pm_LZ_L$ and $W^\pm_Lh$ converge for the following reasons:
\begin{itemize}
\item $G^0$ and $h$ are the two neutral components of the Higgs-doublet.  Since $f\bar{f}'\rightarrow W^\pm_LZ_L$ approaches $f \bar{f}'\rightarrow W^\pm_L G^0$, both  $W^\pm_LZ_L$ and $W^\pm_Lh$ project out the same component of the Higgs doublet.
\item Both processes proceed through an $s$-channel $W^\pm$ exchange. Since both channels have the same initial state quarks, there is no ambiguity about quark couplings.
\end{itemize}

In SMEFT, the operator $\mathcal Q_{Hq}^{(3)}$ modifies the left-handed quark current coupling to electroweak fields after electroweak symmetry breaking. Therefore, $\mathcal{Q}_{Hq}^{(3)}$ shifts the couplings relevant for both $q_-\bar q'_+ \to W^\pm_LZ_L$ and $q_-\bar q'_+ \to W^\pm_Lh$.  The charged-current part of $\mathcal Q_{Hq}^{(3)}$ contains the schematic four-point interaction
\[
\Delta\mathcal L_{\rm contact}^{(4)}
\propto
C_{Hq}^{(3)}\,
i\,(\bar q'_L\gamma^\mu q_L)\,
G^\pm\overleftrightarrow{\partial_\mu}(\pm G^0-i\,h)
+\text{h.c.}\,.
\]
Since $h$ and $G^0$ are the two neutral components of the  Higgs-doublet, this operator induces the same contact term structure in the Goldstone-equivalent amplitudes $q_-\bar q'_+ \to G^\pm G^0$ and $q_-\bar q'_+ \to G^\pm h$. This is why $q_-\bar{q}'_+ \rightarrow W^\pm_LZ_L$ and $q_-\bar{q}'_+\rightarrow W^\pm_L h$ share the same leading $\mathcal O(s)$ behavior in SMEFT.

In contrast to SMEFT, the $\mathcal{O}(s)$ terms in the HEFT amplitudes for $q_-\bar{q}'_+\rightarrow W_L^\pm Z_L$ and $q_-\bar{q}'_+\rightarrow W_L^\pm h$ are controlled by independent combinations of HEFT operators.  For example, $\mathcal{P}_{3}(h)$ contributes energy squared growth to $q_-\bar{q}'_+\rightarrow W^\pm_LZ_L$ but not $q_-\bar{q}'_+\rightarrow W^\pm_L h$. Also, the combination of $n_1^Q+n_7^Q$ contributes to $W^\pm_LZ_L$ and $\kappa_1n_1^Q-\kappa_7n_7^Q$ contributes to $W^\pm_Lh$. Furthermore, the values of $\kappa_1\neq -1$ and $\kappa_7\neq 1$ further modify the ratio. As a result, at linear order in the HEFT expansion, the ratio of longitudinal amplitudes $\mathcal{M}_{\rm HEFT}(q_-\bar{q}'_+\rightarrow W_L^\pm Z_L)$ and $\mathcal{M}_{\rm HEFT}(q_-\bar{q}'_+\rightarrow W^\pm_Lh)$ is sensitive to variations of $n_{1,7}^Q$, $\kappa_{1,7}$ and $c_{3}$. Similar discussion also holds for $q_+\bar{q}'_-\rightarrow W_L^\pm Z_L$ and $q_+\bar{q}'_-\rightarrow W_L^\pm h$.

In the SM, the ratios of the other amplitudes do not converge to one as may be naively expected from GBET. This is because these processes have different initial state quarks, different quark helicities, and/or different $s$-channel propagators. For example, $W^+_LW^-_L$ production has a photon propagator contribution while $Z_Lh$ does not. Hence, the ratios of all the amplitudes are different, although they should converge to a constant value. Similarly, in SMEFT different operators contribute to charged vs. neutral current contact interactions. Hence, the ratios depend on which operators are present. Therefore, similarly to HEFT, the value that the ratios converge to depends on the values of the Wilson coefficients. In the next section, we will identify a set of ratios of amplitudes that do indeed converge to one for the SM and SMEFT, but not necessarily HEFT.

\subsection{Linear vs. Non-linear and the Goldstone Boson Equivalence Theorem}
\label{sec:GBET}
While in Table~\ref{tab:AmpRat} we have shown the ratios of amplitudes via explicit calculations, it is clear that by combining amplitudes, many of the ratios could be simplified. In this subsection, we examine the leading behavior of the amplitudes by analyzing the $SU(2)_L$ triplet and singlet vector currents that govern the Goldstone boson production amplitudes in the SM and SMEFT at high energies. This analysis is used to predict the ratios of amplitudes in Table~\ref{tab:AmpRatEWR}.

First, we discuss the case of the SM and how EW restoration and the GBET informs which di-boson amplitudes are proportional to each other. In the EW restored SM, the pair production of Goldstone and/or Higgs bosons proceed through the $s$-channel exchange of hypercharge and $SU(2)_L$ gauge bosons. The hypercharge gauge boson mediates a coupling between the quark and Higgs $SU(2)_L$ singlet currents:
\begin{eqnarray}
J^\mu_Q J_{H,\mu} \quad J^\mu_{q} J_{H,\mu}\label{eq:SingletProd}
\end{eqnarray}
where the Higgs $SU(2)_L$ singlet current is
\begin{eqnarray}
J_{H,\mu}&=&H^\dagger i\,\overleftrightarrow{\partial}_\mu H \label{eq:HiggsSinglet}
\end{eqnarray}
and left-handed ($J_{Q,\mu}$) and right-handed ($J_{q,\mu}$) quark $SU(2)_L$ singlet currents are:
\begin{eqnarray}
J_{q,\mu}&=&\bar{q}_R\gamma^\mu q_R,\quad J_{Q,\mu}=\bar{Q}_L\gamma^\mu Q_L\label{eq:qSinglet}.
\end{eqnarray}
The $SU(2)_L$ gauge bosons mediate a coupling between the left-handed quark and Higgs $SU(2)_L$ triplet vector currents
\begin{eqnarray}
J^{I,\mu}_Q J^I_{H,\mu},\label{eq:TripletProduct}
\end{eqnarray}
where the Higgs $SU(2)_L$ triplet current is
\begin{eqnarray}
J^I_{H,\mu} =H^\dagger \sigma^I\, i\,\overleftrightarrow{\partial}_\mu H,\label{eq:HiggsIsospinTriplet}
\end{eqnarray}
and the left-handed quark $SU(2)_L$ triplet current is
\begin{eqnarray}
J^I_{Q,\mu}=\bar{Q}_L\gamma_\mu \sigma^I Q_L.\label{eq:QLIsospinTriplet}
\end{eqnarray}

The products of currents in Eqs.~(\ref{eq:SingletProd},\ref{eq:TripletProduct}) can be used to derive couplings between the electric charge eigenstate quarks, Goldstone bosons, and Higgs bosons in the broken phase of the EW theory.
The product of $SU(2)_L$ singlet currents in Eq.~(\ref{eq:SingletProd})  have the form
\begin{subequations}
\begin{eqnarray}
J^\mu_QJ_{H,\mu}\supset i\,\bar{q}_L\gamma^\mu q_L\left[G^-\overleftrightarrow{\partial}_\mu G^+-i\,G^0\overleftrightarrow {\partial}_\mu h\right],\label{eq:LeftSinglet}\\
J^\mu_qJ_{H,\mu}\supset i\,\bar{q}_R\gamma^\mu q_R\left[G^-\overleftrightarrow{\partial}_\mu G^+-i\,G^0\overleftrightarrow{\partial}_\mu h\right]\label{eq:RightSinglet},
\end{eqnarray}
\end{subequations}
where $q$ can be up or down quarks.
The $SU(2)_L$ triplet currents can be rewritten into charged ($J^\pm$) and neutral ($J^0$) currents:
\begin{eqnarray}
J^\pm_\mu &=& \frac{1}{\sqrt{2}}\left(J_\mu^1\pm i\,J^2_\mu\right),\quad J^0_\mu = J_\mu^3,
\end{eqnarray}
The following products for charged and neutral currents make contributions to Eq.~(\ref{eq:TripletProduct}):
\begin{eqnarray}
J^{I,\mu}_Q J^I_{H,\mu}&=&J^{+,\mu}_{Q}J^{-}_{H,\mu}+J^{-,\mu}_{Q}J^{+}_{H,\mu}+J^{0,\mu}_{Q}J^{0}_{H,\mu}\\
J^{\mp,\mu}_{Q}J^{\pm}_{H,\mu}&\supset&i\,\sqrt{2}\,\bar{q}'_L\gamma^\mu q_L\left[\mp G^\mp \overleftrightarrow{\partial}_\mu G^0+i\,G^\mp\overleftrightarrow{\partial}_\mu h\right]\nonumber\\
J^{0,\mu}_{Q}J^{0}_{H,\mu}&\supset&i\,2\,T_3^q\bar{q}_L\gamma^\mu q_L \left[G^-\overleftrightarrow{\partial}_\mu G^++i\,G^0\overleftrightarrow{\partial}_\mu h\right].\nonumber
\end{eqnarray}

In the high energy and massless quark limit, the products of currents contribute to the following combinations of Goldstone and Higgs amplitudes with fixed relative sign and magnitude: 
\begin{subequations}
\begin{eqnarray}
J_q^\mu J_{H,\mu}:& \mathcal{M}_{\rm SM}(q_+\bar{q}_-\rightarrow G^+(p_1)G^-(p_2))\label{eq:qRNeutSing}\\
&~-i\mathcal{M}_{\rm SM}(q_+\bar{q}_-\rightarrow G^0(p_1)h(p_2))\nonumber\\
J_Q^\mu J_{H,\mu}:& \mathcal{M}_{\rm SM}(q_-\bar{q}_+\rightarrow G^+(p_1)G^-(p_2))\label{eq:qLNeutSing}\\
&~-i\mathcal{M}_{\rm SM}(q_-\bar{q}_+\rightarrow G^0(p_1)h(p_2))\nonumber\\
J^{\mp,\mu}_QJ^{\pm}_{H,\mu}: & \mp\sqrt{2}\mathcal{M}_{\rm SM}(q_-\bar{q}'_+\rightarrow G^\pm(p_1)G^0(p_2))\label{eq:qLChargeTrip}\\
&~+i\sqrt{2}\mathcal{M}_{\rm SM}(q_-\bar{q}'_+\rightarrow G^\pm(p_1)h(p_2))\nonumber\\
J_Q^{0,\mu} J^0_{H,\mu}:& ~~2\,T_3^q\left[\mathcal{M}_{\rm SM}(q_-\bar{q}_+\rightarrow G^+(p_1)G^-(p_2))\right.\label{eq:qLNeutTrip}\\
&\left.~+i\mathcal{M}_{\rm SM}(q_-\bar{q}_+\rightarrow G^0(p_1)h(p_2))\right]\nonumber
\end{eqnarray}
\end{subequations}

In the SM, since right handed quark initial states only have contributions from the $SU(2)_L$ singlet current, Eq.~(\ref{eq:qRNeutSing}) can be used to infer the relationship 
\begin{eqnarray}
&\displaystyle\mathcal{M}_{\rm SM}&(q_+\bar{q}_-\rightarrow G^+(p_1)G^-(p_2))\label{eq:qRSMGolds}\\
&&\displaystyle\quad=-i\,\mathcal{M}_{\rm SM}(q_+\bar{q}_-\rightarrow G^0(p_1)h(p_2))\left[1+\mathcal{O}\left(\frac{m}{\sqrt{s}}\right)\right]\nonumber
\end{eqnarray}
For left-handed initial state quarks, 
the $SU(2)_L$ singlet and triplet currents contribute to independent linear combinations of matrix elements with neutral final states in Eqs.~(\ref{eq:qLNeutSing},\ref{eq:qLNeutTrip}). Together with the fact that only the $SU(2)_L$ triplet currents contribute to charged final states, we can infer the following relations from Eqs~(\ref{eq:qLChargeTrip},\ref{eq:qLNeutTrip}):
\begin{widetext}
\begin{eqnarray}
\mp\sqrt{2}\mathcal{M}_{\rm SM}(q_-\bar{q}'_+\rightarrow G^\pm (p_1)G^0(p_2))&=&\sqrt{2}i\,\mathcal{M}_{\rm SM}(q_-\bar{q}'_+\rightarrow G^\pm(p_1)h(p_2))\left[1+\mathcal{O}\left(\frac{m}{\sqrt{s}}\right)\right]\label{eq:qLSMGolds}\\
&=&2\,T_3^q\left[\mathcal{M}_{\rm SM}(q_-\bar{q}_+\rightarrow G^+(p_1)G^-(p_2))+i\mathcal{M}_{\rm SM}(q_-\bar{q}_+\rightarrow G^0(p_1)h(p_2)\right]\left[1+\mathcal{O}\left(\frac{m}{\sqrt{s}}\right)\right]\nonumber.
\end{eqnarray}
\end{widetext}

The relationships among the Goldstone amplitudes in Eqs.~(\ref{eq:qRSMGolds},\ref{eq:qLSMGolds}) can be used to find relationships among amplitudes with outgoing longitudinal gauge bosons using the GBET~\cite{Chanowitz:1985hj,Yao:1988aj,Bagger:1989fc,He:1992nga}: 
\begin{eqnarray}
&&\mathcal{M}(V_L^1,\ldots,V_L^n,{\rm physical~fields})=\\
&&\quad\quad(i)^n\mathcal{M}(\phi^1,\ldots,\phi^n,{\rm physical~fields})+\mathcal{O}\left(\frac{m}{\sqrt{s}}\right).\nonumber
\end{eqnarray}
From Eqs.~(\ref{eq:qRSMGolds},\ref{eq:qLSMGolds}), we get the high energy relationships among longitudinal gauge boson final states:
\begin{widetext}
\begin{eqnarray}
\displaystyle\mathcal{M}_{\rm SM}(q_+\bar{q}_-\rightarrow W^+_L(p_1)W^-_L(p_2))&=&\mathcal{M}_{\rm SM}(q_+\bar{q}_-\rightarrow Z_L(p_1)h(p_2))\left[1+\mathcal{O}\left(\frac{m}{\sqrt{s}}\right)\right]\label{eq:SMEWR}\\
\pm\sqrt{2}\mathcal{M}_{\rm SM}(q_-\bar{q}_+'\rightarrow W^\pm_L(p_1)Z_L(p_2))&=&\sqrt{2}\mathcal{M}_{\rm SM}(q_-\bar{q}'_+\rightarrow W^\pm_L(p_1)h(p_2))\left[1+\mathcal{O}\left(\frac{m}{\sqrt{s}}\right)\right]\nonumber\\
\quad &=&2\,T_3^q\left[\mathcal{M}_{\rm SM}(q_-\bar{q}_+\rightarrow Z_L(p_1)h(p_2))-\mathcal{M}_{\rm SM}(q_-\bar{q}_+\rightarrow W^+_L(p_1)W^-_L(p_2))\right]\left[1+\mathcal{O}\left(\frac{m}{\sqrt{s}}\right)\right]\nonumber
\end{eqnarray}
\end{widetext}
These relations are those shown in the first four rows of Table~\ref{tab:AmpRatEWR}.

As mentioned above, for left-handed initial state quarks, the $s$-channel hypercharge and neutral $SU(2)_L$ gauge boson exchange contribute to two independent linear combinations of $G^+G^-$ and $G^0h$ amplitudes in Eqs.~(\ref{eq:qLNeutSing},\ref{eq:qLNeutTrip}). Since these combinations come from different force mediators, we expect the ratio to be proportional to the ratio of the corresponding gauge couplings squared and the couplings of the initial state quarks to neutral $SU(2)_L$ and hypercharge gauge boson (up to an overall sign). Using the GBET, in the SM this would predict
\begin{widetext}
\begin{eqnarray}
\frac{\mathcal{M}_{\rm SM}(q_-\bar{q}_+\rightarrow Z_L(p_1)h(p_2))-\mathcal{M}_{\rm SM}(q_-\bar{q}_+\rightarrow W^+_L(p_1)W^-_L(p_2))}{\mathcal{M}_{\rm SM}(q_-\bar{q}_+\rightarrow Z_L(p_1)h(p_2))+\mathcal{M}_{\rm SM}(q_-\bar{q}_+\rightarrow W^+_L(p_1)W^-_L(p_2))}=\pm \frac{g^2T_3^q}{{g'}^2Y_L^q}+\mathcal{O}\left(\frac{m}{\sqrt{s}}\right)\label{eq:SingTripRat}.
\end{eqnarray}
\end{widetext}
Again, this is precisely the SM relation seen in Table~\ref{tab:AmpRatEWR}.

In SMEFT, there are $\mathcal{O}(s)$ contributions to the di-boson amplitudes from the operators in Eqs.~(\ref{eq:QHq}-\ref{eq:QHud}). The operators $\mathcal{Q}_{Hq},\, \mathcal{Q}_{Hq}^{(1)},\,\mathcal{Q}_{Hq}^{(3)}$ follow the same pattern of products of quark and Higgs currents as the SM. Hence, SMEFT amplitudes with longitudinal gauge bosons will also obey the relations in Eq.~(\ref{eq:SMEWR}). SMEFT also has a similar relationship as Eq.~(\ref{eq:SingTripRat}), however, now the ratio is between the Wilson coefficients of the product of singlet currents, $\mathcal{Q}_{Hq}^{(1)}$, and triplet currents, $\mathcal{Q}_{Hq}^{(3)}$:
\begin{widetext}
\begin{eqnarray}
\frac{\mathcal{M}_{\rm SMEFT}(q_-\bar{q}_+\rightarrow Z_L(p_1)h(p_2))-\mathcal{M}_{\rm SMEFT}(q_-\bar{q}_+\rightarrow W^+_L(p_1)W^-_L(p_2))}{\mathcal{M}_{\rm SMEFT}(q_-\bar{q}_+\rightarrow Z_L(p_1)h(p_2))+\mathcal{M}_{\rm SMEFT}(q_-\bar{q}_+\rightarrow W^+_L(p_1)W^-_L(p_2))}=\pm 2\,T_3^q\frac{C_{Hq}^{(3)}}{C_{Hq}^{(1)}}+\mathcal{O}\left(\frac{m}{\sqrt{s}}\right)\label{eq:SingTripRat}.
\end{eqnarray}
\end{widetext}
where the quark isospin is included. 
In addition to the SM relations discussed above, SMEFT has an additional relationship coming from $\mathcal{Q}_{Hud}$. This amplitude couples right-handed fermions to the Higgs current
\begin{eqnarray}
\widetilde{J}_\mu &=&\widetilde{H}^\dagger i\overleftrightarrow{\partial}_\mu H.
\end{eqnarray}
This is a charged current:
\begin{eqnarray}
\widetilde{J}^\pm_\mu&\supset& \sqrt{2}i\,G^\mp\overleftrightarrow{\partial}_\mu(i\,h\pm G^0).
\end{eqnarray}
Following the same arguments as above, this leads to a relationship between right-handed initial state quarks and charged final states
\begin{eqnarray}
&&\mathcal{M}_{\rm SMEFT}(q_+\bar{q}'_-\rightarrow W^\pm_L(p_1)h(p_2))\\
&&\quad =\mp \mathcal{M}_{\rm SMEFT}(q_+\bar{q}_-'\rightarrow W_L^\pm(p_1)Z_L(p_2))\left[1+\mathcal{O}\left(\frac{m}{\sqrt{s}}\right)\right].\nonumber
\end{eqnarray}
This discussion explains the ratios of amplitudes in SMEFT as shown in Table~\ref{tab:AmpRatEWR}.

As the above discussion shows, in the linear EW symmetry, the GBET and EW restoration can be used to derive a set of proportionality relationships of amplitudes with longitudinal gauge bosons at high energy. These relationships are verified by the explicit calculations for SM and SMEFT amplitudes in Sec.~\ref{sec:theory}. As is clear from those calculations and Tab.~\ref{tab:AmpRatEWR}, the HEFT amplitudes do not obey the same relations. This is due to the Higgs boson being a singlet and custodial symmetry breaking effects appearing at $\mathcal{O}(s)$. 

The ratios in the first four rows of Table~\ref{tab:AmpRatEWR} provide a robust test of linear vs. non-linear EW symmetry since in SMEFT they are independent of BSM parameters while in HEFT they depend on BSM parameters. While the ratio in the last row of Table~\ref{tab:AmpRatEWR} can be derived in the SM and SMEFT, it is not as useful to distinguish linear and non-linear EW symmetry since both in SMEFT and HEFT this ratio depends on undetermined BSM parameters.   

In HEFT, the Goldstone bosons form an $SU(2)_L$ triplet. Hence, we still expect there to be relationships between the $W^+_LW^-_L$ and $W^\pm_L Z_L$ amplitudes. For example, from the explicit calculations we find:
\begin{widetext}
\begin{subequations}
\begin{eqnarray}
\mathcal{M}_{\rm HEFT}(q_-\bar{q}'_+\rightarrow W^+_LW^-_L)&=&\left(\mp \sqrt{2}T_3^q \mathcal{M}_{\rm HEFT}(q_-\bar{q}'_+\rightarrow W^\pm_L Z_L)+4\sqrt{2} {G}_Fs\left(n_5^Q+4\,T_3^q\,\,n_7^Q\right)\,\sin\theta\right)\nonumber\\
&&\hspace{18em}\times\left[1+\mathcal{O}\left(\frac{m^2}{s}\right)\right].\\
\mathcal{M}_{\rm HEFT}(q_+\bar{q}'_-\rightarrow W^+_LW^-_L)&=&\left(\mp\sqrt{2}T_3^q \mathcal{M}_{\rm HEFT}(q_+\bar{q}'_-\rightarrow W^\pm_L Z_L)-4\sqrt{2}\, {G}_Fs\left(n_6^Q+2\,T_3^q\,n_8^Q\pm 2\,i\,T_3^q\,n_4^Q\right)\,\sin\theta\right)\nonumber\\
&&\hspace{18em}\times\left[1+\mathcal{O}\left(\frac{m^2}{s}\right)\right].
\end{eqnarray}
\end{subequations}
\end{widetext}
That is, for the subset of operators considered here, the proportionality of the $W_L^+W_L^-$ and $W_L^\pm Z_L$ amplitudes at high energies is broken by the custodial symmetry violating operators $\mathcal{N}_{4,5,6,7,8}^Q$.

Additionally, there is a limit in HEFT that will reproduce to SMEFT couplings. For example, from the explicit calculations in the third and fourth row of Tab.~\ref{tab:AmpRatEWR}, choosing $\kappa_2=\kappa_6=\kappa_8=1$ will reproduce the leading SM and SMEFT results at high energy.  Indeed, the leading high energy $\mathcal{O}(s)$ behavior of the SMEFT amplitudes for the longitudinal di-boson production calculated in Sec.~\ref{sec:theory} can be fully reproduced with the following choices of parameters in HEFT:
\begin{subequations}
\begin{eqnarray}
c_3&=&\frac{8\pi}{ {g}}\left(1-\kappa_1\right)n_1^Q\\
\kappa_1n_1^Q&=&\frac{C_{Hq}^{(3)}}{2\sqrt{2} {G}_F}\\
n_2^Q&=&\frac{C_{Hd}-C_{Hu}+{\rm Re}(C_{Hud})}{8\sqrt{2} {G}_F}\\
n_4^Q&=&\frac{{\rm Im}(C_{Hud})}{8\sqrt{2} {G}_F}\\
n_5^Q&=&-\frac{C_{Hq}^{(1)}}{4\sqrt{2} {G}_F}\\
n_6^Q&=&-\frac{C_{Hd}+C_{Hu}}{8\,\sqrt{2}\, {G}_F}\\
n_7^Q&=&0\\
n_8^Q&=&\frac{C_{Hd}-C_{Hu}-{\rm Re}(C_{Hud})}{8\sqrt{2} {G}_F}\\
\kappa_2&=&\kappa_4=\kappa_5=\kappa_6=\kappa_8=1
\end{eqnarray}
\end{subequations}
This implies the following relationships among the HEFT Wilson coefficients that would reproduce SMEFT-like behavior at $\mathcal{O}(s)$ for the set of amplitudes $q\bar{q}'\rightarrow W^+_LW^-_L/W^\pm_LZ_L/W^\pm_Lh/Z_Lh$:
\begin{subequations}
\begin{eqnarray}
c_3&=&\frac{8\pi}{ {g}}(1 -\kappa_1)n_1^Q\\
n_7^Q&=&0\\
\kappa_2&=&\kappa_4=\kappa_5=\kappa_6=\kappa_8=1,
\end{eqnarray}
\end{subequations}
with all other Wilson coefficients free.  

Of course, there are many other HEFT operators~\cite{Brivio:2016fzo} that can contribute to these processes beyond those that we considered in Eqs.~(\ref{eq:HEFTn1Q}-\ref{eq:HEFTp3}). Including more operators would introduce more restrictions to reproduce the SMEFT predictions and could alleviate the $n_7^Q=0$ constraint. However, all relevant operators in dimension-6 SMEFT have been included, and even with this subset of HEFT operators, it is clear from Table~\ref{tab:AmpRatEWR} that the SM and dimension-6 SMEFT with linear EW symmetry realizations have predictions for the high energy behavior of amplitudes that do not generically hold in the non-linear HEFT.

\subsection{Partonic Cross Sections}

\begin{figure*}
\includegraphics[width=0.43\textwidth]{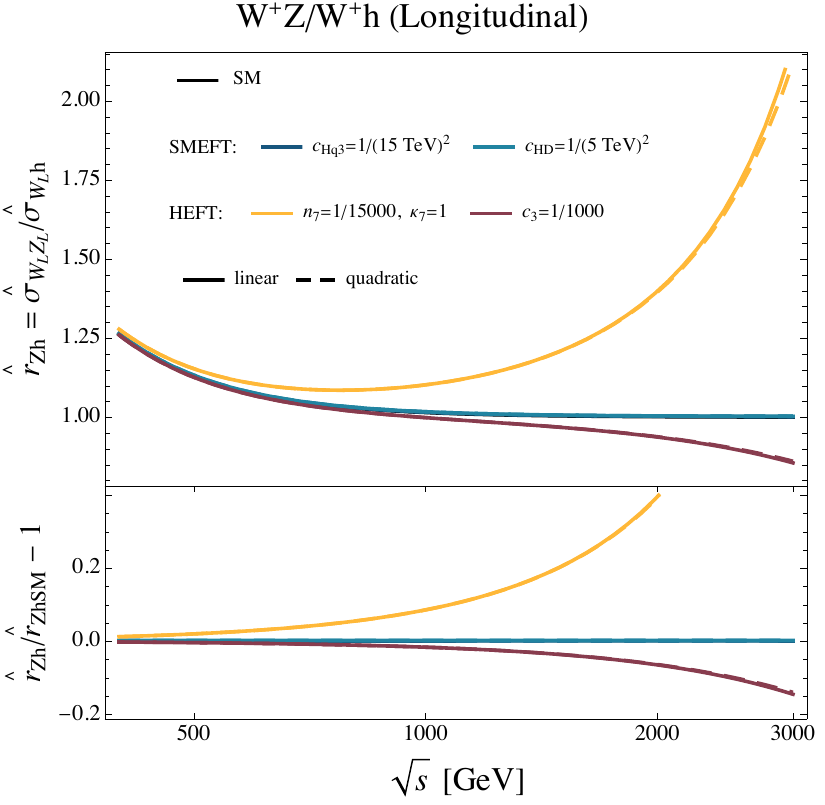}
\includegraphics[width=0.43\textwidth]{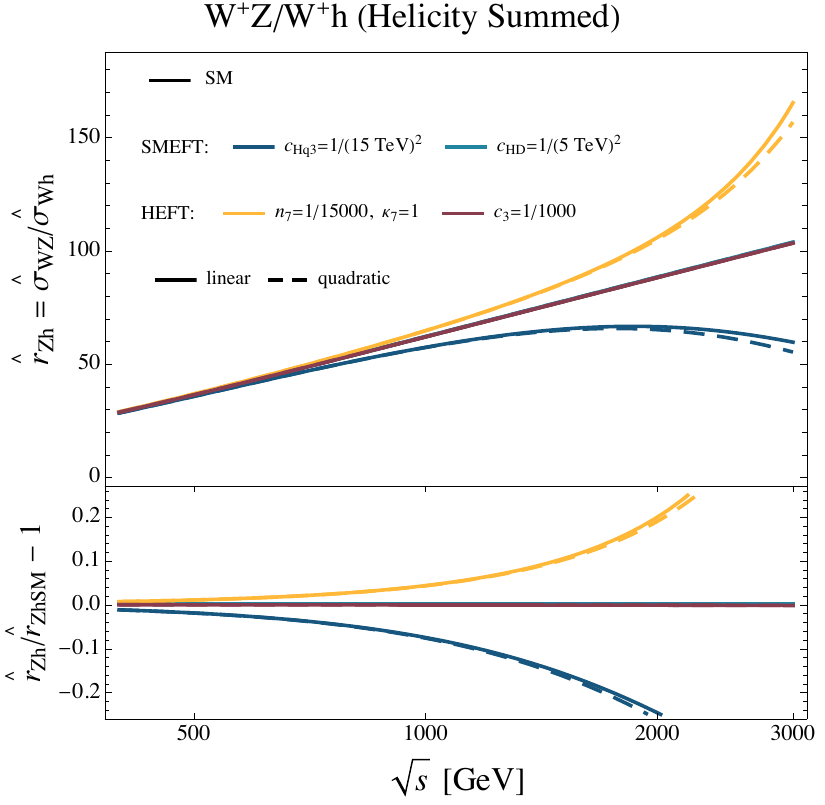}\\

\includegraphics[width=0.43\textwidth]{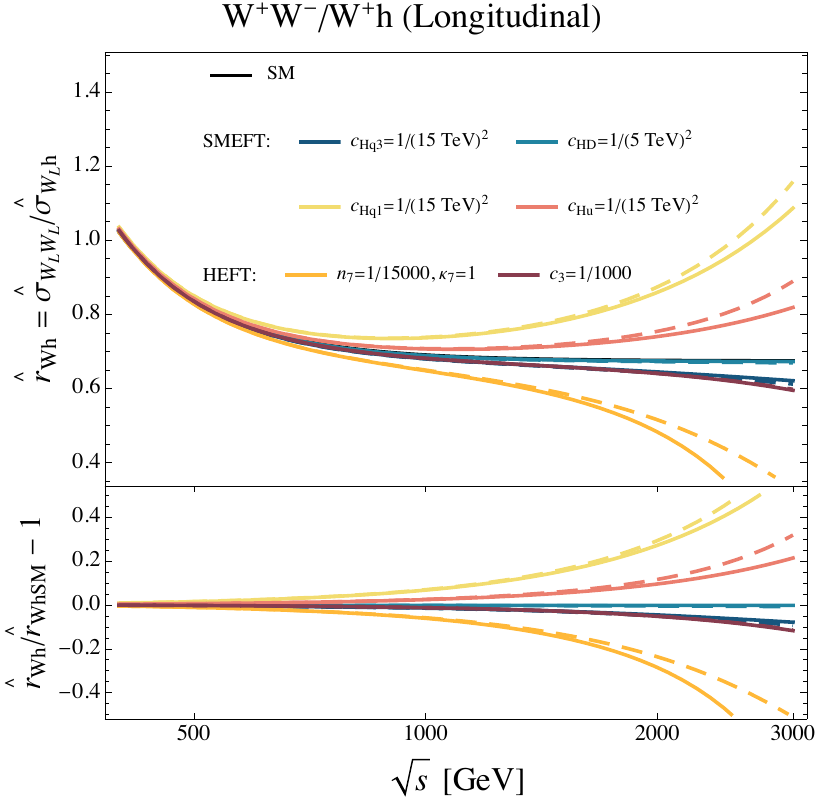}
\includegraphics[width=0.43\textwidth]{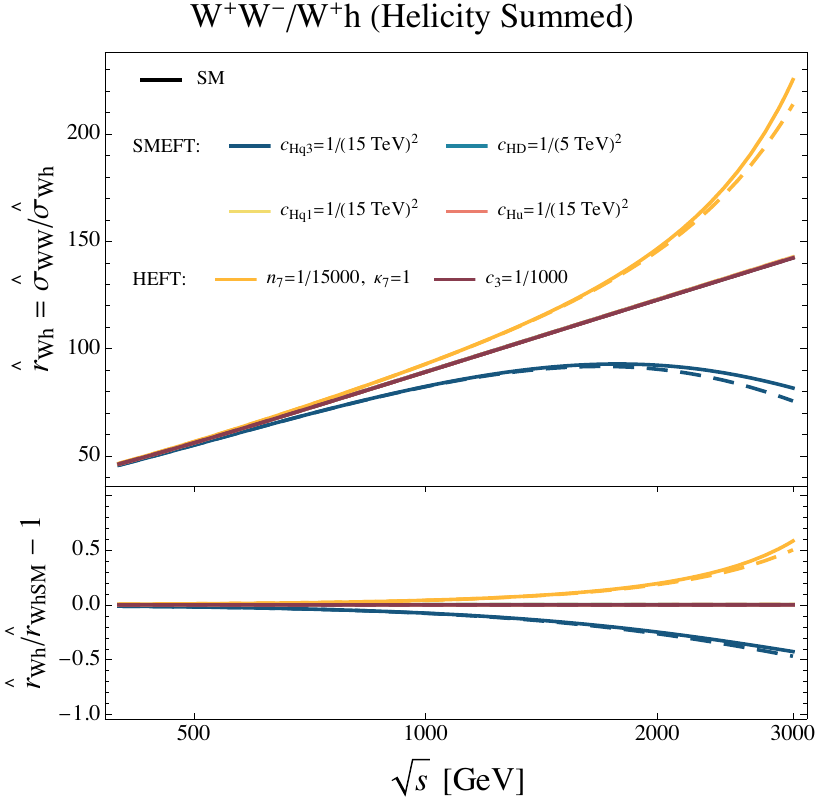}\\

\includegraphics[width=0.43\textwidth]{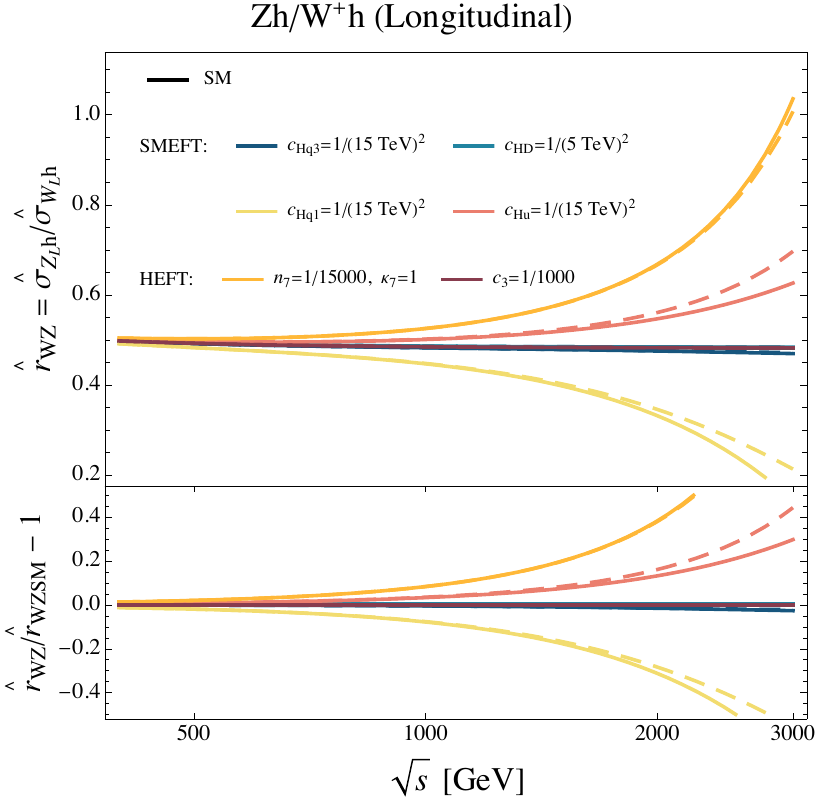}
\includegraphics[width=0.43\textwidth]{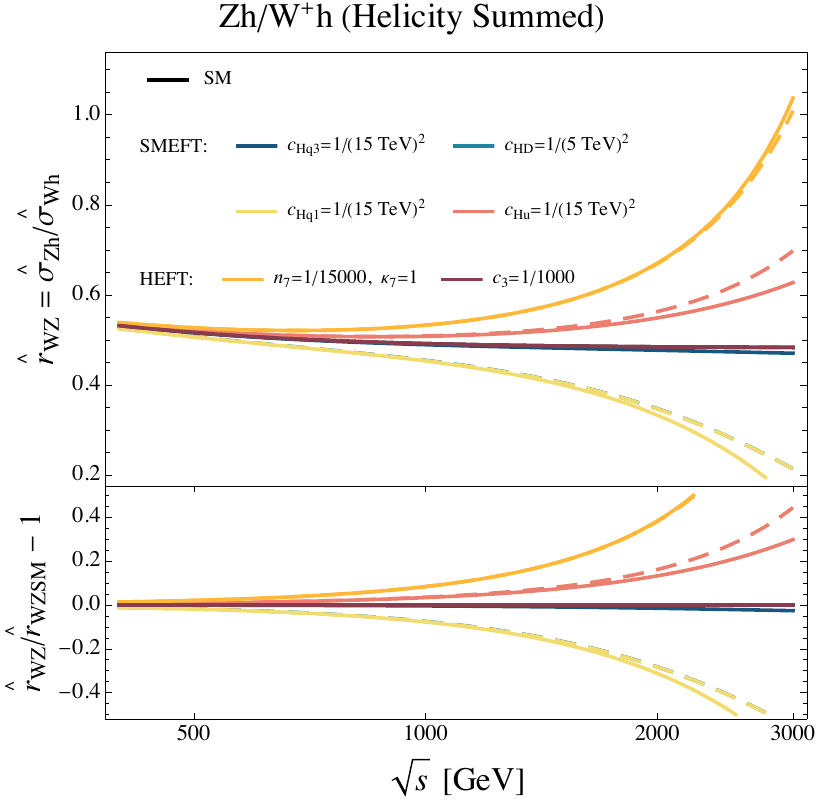}

\caption{\label{fig:xsectRat}   Ratios of partonic cross sections as a function of partonic center of momentum energy for initial state up-type quarks: (top) $\hat{r}_{Zh}$, (middle) $\hat{r}_{Wh}$, and (bottom) $\hat{r}_{WZ}$ as defined in Eqs.~(\ref{eq:rZh}-\ref{eq:rWZ}).  
Plots in the left panel have fully longitudinal final state gauge bosons, while in the right panel, the gauge boson helicities are summed. 
In all plots we show (black) SM, (dark blue, yellow, red) $C_{Hq}^{(3)}=C_{Hq}^{(1)}=C_{Hu}=(15~{\rm TeV})^{-2}$, (light blue) $C_{HD}=(5~{\rm TeV})^{-2}$, (orange) $n_7=1/15000$, and (maroon) $c_{3}=10^{-3}$. Solid curves include linear EFT contributions, while dashed curves include quadratic EFT contributions.}
\end{figure*}

Now we discuss the ratios of partonic cross sections. To keep our calculations valid in the EFT, we choose parameter points that are consistent with current constraints~\cite{deBlas:2025xhe} and where the quadratic contribution to the cross section from the Wilson coefficients is at most 10\% of the linear contribution at partonic center of momentum energies of $\sqrt{s}=3$~TeV.  For our numerical analysis, we only choose a subset of HEFT operators to illustrate how they can break the SMEFT expectations. Additionally, we do not include $\mathcal{Q}_{Hud}$ since by minimal flavor violation~\cite{Chivukula:1987py,DAmbrosio:2002vsn} it would be expected to be suppressed by light quark masses. Our parameter points are chosen as:
\begin{eqnarray}
&{\rm SMEFT}:\quad &
C_{Hq}^{(3)}=C_{Hq}^{(1)}=\,C_{Hu}=\left(15~{\rm TeV}\right)^{-2}\label{eq:params}\\\nonumber
&&C_{HD}=\left(5~{\rm TeV}\right)^{-2}\nonumber\\
&{\rm HEFT}:&c_{3}=10^{-3}, \,n_7^Q=\frac{2}{3}\times 10^{-4},\, \kappa_7 =1\nonumber 
\end{eqnarray}
Only one operator is considered at a time, with all other Wilson coefficients set to zero.

In Fig.~\ref{fig:xsectRat} we show the ratios of various partonic cross sections with initial state quark helicities summed:
\begin{subequations}
\begin{eqnarray}
 \hat{r}_{Zh}&=&\frac{ {\sigma}(u\bar{d}\rightarrow W^+Z)}{ {\sigma}(u\bar{d}\rightarrow W^+h)}\label{eq:rZh}\\
 \hat{r}_{Wh}&=&\frac{ {\sigma}(u\bar{u}\rightarrow W^+W^-)}{ {\sigma}(u\bar{d}\rightarrow W^+h)}\label{eq:rWh}\\
 \hat{r}_{WZ}&=&\frac{ {\sigma}(u\bar{u}\rightarrow Zh)}{ {\sigma}(u\bar{d}\rightarrow W^+h)}\label{eq:rWZ}.
\end{eqnarray}
\end{subequations}
These ratios are similar to those defined in~\cite{Capdevilla:2024amg}, and we have specialized to up-type quark initial states as they are dominant at the LHC and for illustration purposes.

The left panel of Fig.~\ref{fig:xsectRat} shows purely longitudinal gauge bosons, while on the right panel, the gauge boson helicities are summed. Cross sections with only (solid) linear or (dashed) up to quadratic order in Wilson coefficients are also shown. As can be seen, even though some ratios are getting large, the EFT expansion is still valid in this energy regime for the benchmark Wilson coefficients considered, as the quadratic EFT corrections remain small compared to the linear contributions.

The ratios of $u\bar{d}\rightarrow W^+Z$ and $u\bar{d}\rightarrow W^+h$ are shown in the top two plots of Fig.~\ref{fig:xsectRat}. As expected from the discussion of amplitudes, for longitudinal gauge bosons both the SM and SMEFT ratios converge to one. This is because the leading SM and SMEFT terms in high energy are the same for both amplitudes, hence the cross sections converge\footnote{ Although we neglected $\mathcal{Q}_{Hud}$ here, from Table~\ref{tab:AmpRatEWR} it would also predict $\hat{r}_{Zh}\rightarrow 1$ in SMEFT for longitudinal gauge bosons}. In HEFT, the ratio of cross sections significantly deviates from SM/SMEFT predictions at high energies for both operators $\mathcal{N}_7^Q$ and $\mathcal{P}_{3}$. 
This divergence is particularly striking for the $\mathcal{N}_7^Q$ operator. This can be understood by noting that the SM and $n_7^Q$ contributions to the amplitudes in Eqs.~(\ref{eq:HEFTWZ},\ref{eq:HEFTWh}) have opposite signs when $\kappa_7>0$. Hence, for positive $n_7^Q$, there is energy growing constructive interference in $W_L^+Z_L$ and energy growing destructive interference in $W_L^+h$ production with $\kappa_7=1$.  Hence, the ratio has a striking deviation from SM/SMEFT predictions for the operator $\mathcal{N}_7^Q$. Indeed, the only way to obtain the same energy growth in $W^\pm_LZ_L$ and $W^\pm_Lh$ for $\mathcal{N}_7^Q$ is to fine-tune $\kappa_7=-1$ such that the energy growing interference between HEFT and the SM is either constructive or destructive in both processes. SMEFT has the same energy growth in the two processes without fine-tuning. The operator $\mathcal{P}_{3}$ only has energy growing contributions to $W^+_LZ_L$ and hence causes an energy growing divergence in the ratio as well.

The same conclusion holds for the charge-conjugate processes
\(d\bar u\to W^-Z\) and \(d\bar u\to W^-h\). In SMEFT, the leading
\(\mathcal Q_{Hq}^{(3)}\) contribution has the same relative sign with
respect to the SM amplitude in both \(W^\pm Z\) and \(W^\pm h\), so the
partonic longitudinal ratio \(\hat r_{Zh}\) is unchanged under
\(W^+\leftrightarrow W^-\).

From the top right plot of Fig.~\ref{fig:xsectRat}, the importance of tagging gauge boson polarizations in $W^+Z$ and $W^+h$ is clear. When gauge-boson polarizations are summed, the helicity-summed
\(W^+Z\) and \(W^+h\) ratio no longer shows the same SMEFT--HEFT separation. We see that both the SMEFT operator $\mathcal Q_{Hq}^{(3)}$ and the HEFT operator $\mathcal{N}_7^Q$ generate
large deviations in the ratio, while \(\mathcal Q_{HD}\) and
\(\mathcal P_3\) remain close to the SM prediction. This behavior can be
traced to the fact that the helicity-summed \(u\bar d\to W^+Z\) rate is
dominated by transverse-polarization contributions, and the operators considered here do not generate \(\mathcal O(s)\)
growth in the transverse amplitudes. Therefore, for the benchmark points chosen, the longitudinal EFT corrections do not significantly modify the helicity-summed \(W^+Z\) rate. On the other hand,
the transverse contribution to \(\sigma(u\bar d\to W^+h)\) is suppressed at high energy, so the \(\mathcal O(s)\) corrections to the longitudinal
\(W^+h\) amplitude can provide significant corrections to the cross section. Therefore, the transverse contribution in \(W^+Z\) cross section hides the pattern present in the longitudinal ratio.

The middle and bottom rows of Fig.~\ref{fig:xsectRat} show the other two independent cross section ratios $ \hat{r}_{Wh}$ and $ \hat{r}_{WZ}$.  While
the SM ratios still converge to constants dictated by the initial-state quark
couplings, even when the final states are restricted to longitudinally
polarized gauge bosons, there are parameter choices for which both SMEFT and
HEFT deviate from the SM predictions with increasing energy. The different
behavior of \(\hat r_{Zh}\), compared with \(\hat r_{Wh}\) and \(\hat r_{WZ}\), arises because the latter ratios involve final states whose
energy growing SMEFT contributions depend on different combinations and relative signs of the Wilson coefficients.   For example, $\mathcal{Q}_{Hu}$ and $\mathcal{Q}_{Hq}^{(1)}$ can only contribute to $\mathcal{O}(s)$ growth in the neutral current channels. Hence, when  $C_{Hu}, C_{Hq}^{(1)}\neq 0$, the ratios between charged-current
and neutral-current processes can grow with energy.  Similarly, the relative signs between $C_{Hq}^{(1)}$ and $C_{Hq}^{(3)}$ for the $\mathcal{O}(s)$ growth in the neutral current channel is different depending on the initial-state quark flavor and on the final state being produced. Hence, whether $C_{Hq}^{(1)},\,C_{Hq}^{(3)}$ contribute to constructive or destructive interference with the SM depends on the final state. As such, seeing energy growth in $ \hat{r}_{Wh}$ or $ \hat{r}_{WZ}$ would not be a definitive test between the linear and non-linear realizations of EW symmetry, even when limiting the final state to longitudinally polarized gauge bosons. 

It is worth noting that the situation would change completely if one gains some control of the initial state polarization, e.g., in future lepton colliders, such as the high energy muon colliders. One can then define various asymmetry observables to extract corresponding cross sections and further enhance the tests highlighted in this study.  In particular, the fermion helicity dependent ratios in Table~\ref{tab:AmpRatEWR} could in principle be more completely tested.

\section{Testing Linear vs. Non-Linear Realizations at the LHC: $WZ$ vs. $Wh$}
\label{sec:LHCresults}

\begin{figure*}
\includegraphics[width=0.45\textwidth]{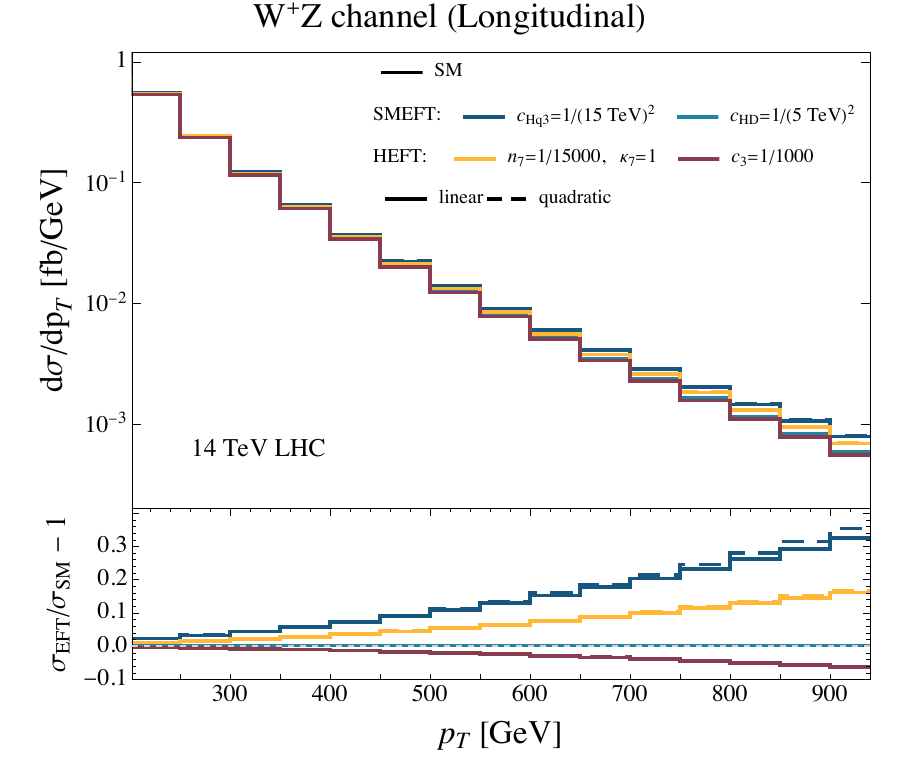}
\includegraphics[width=0.45\textwidth]{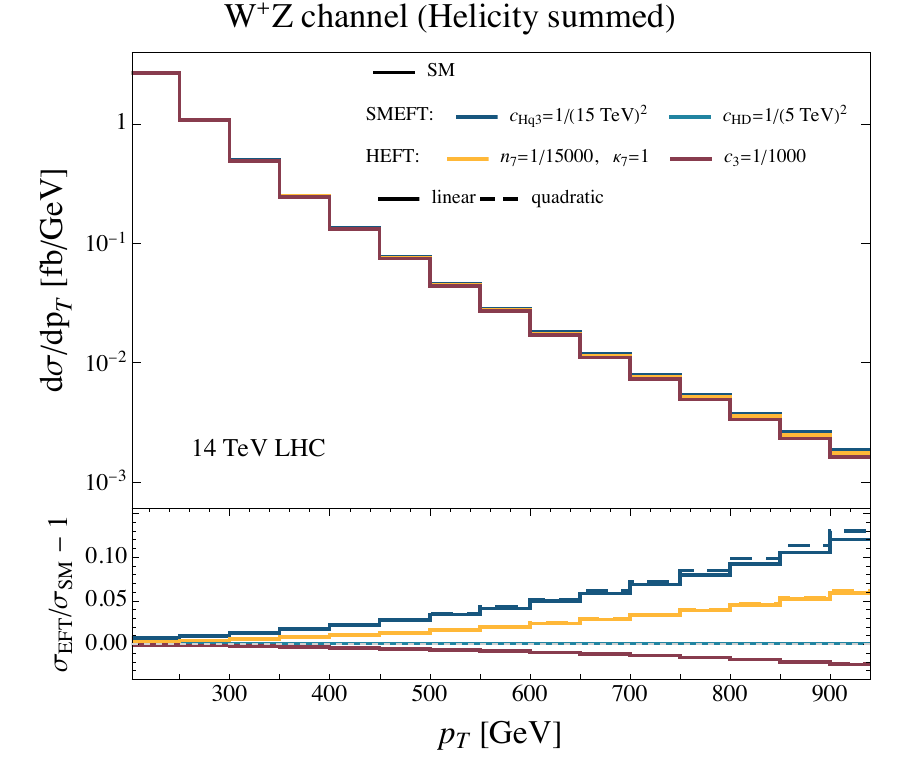}\\

\includegraphics[width=0.45\textwidth]{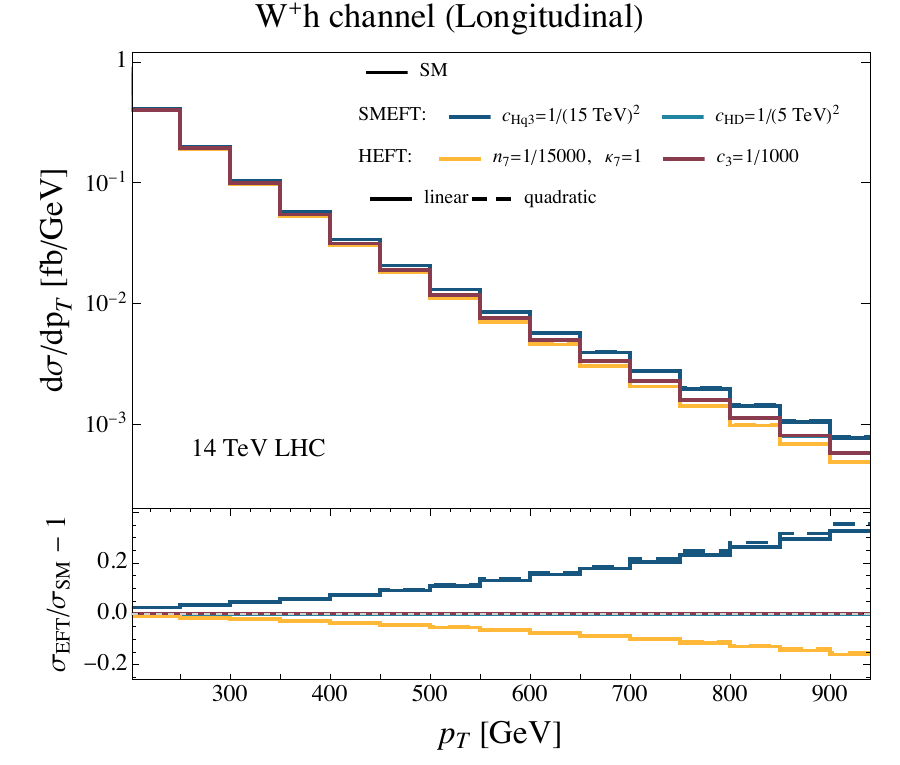}
\includegraphics[width=0.45\textwidth]{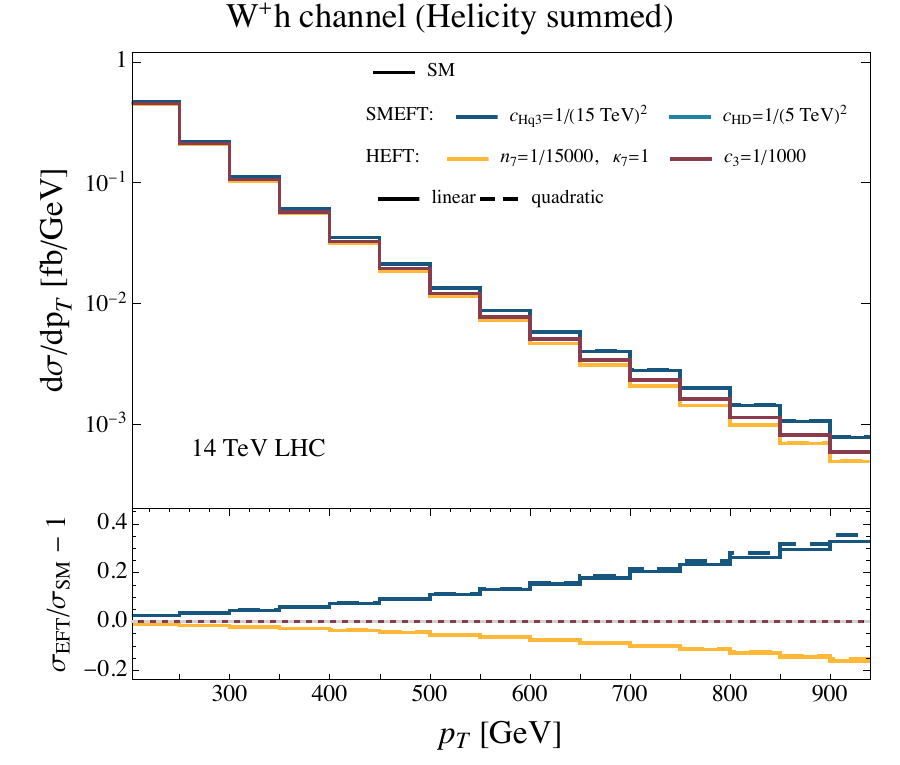}\\

\includegraphics[width=0.45\textwidth]{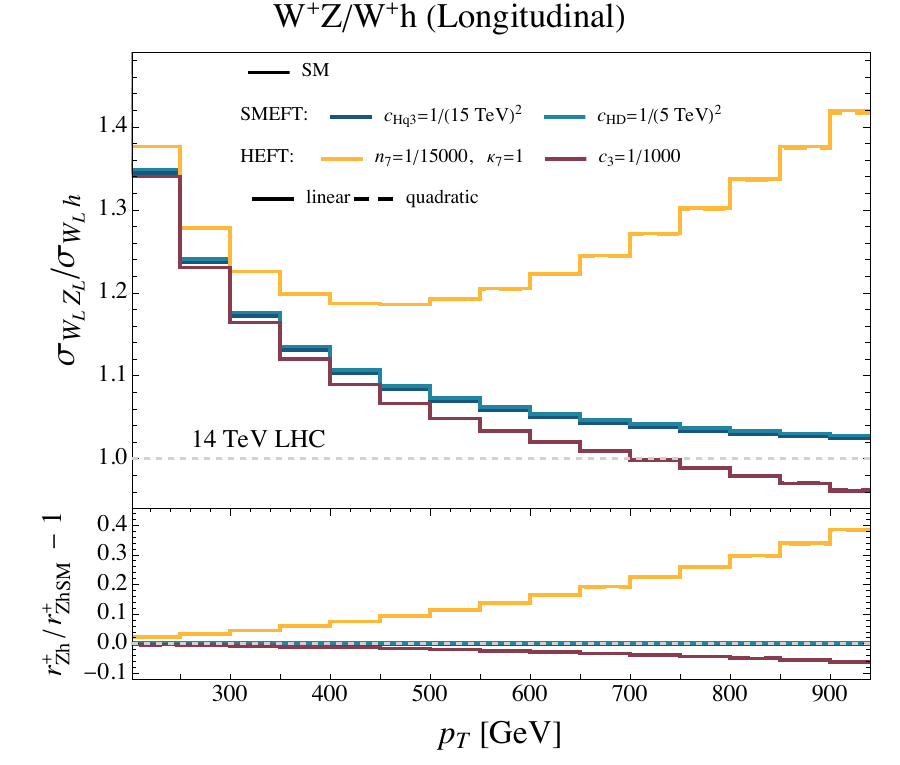}
\includegraphics[width=0.45\textwidth]{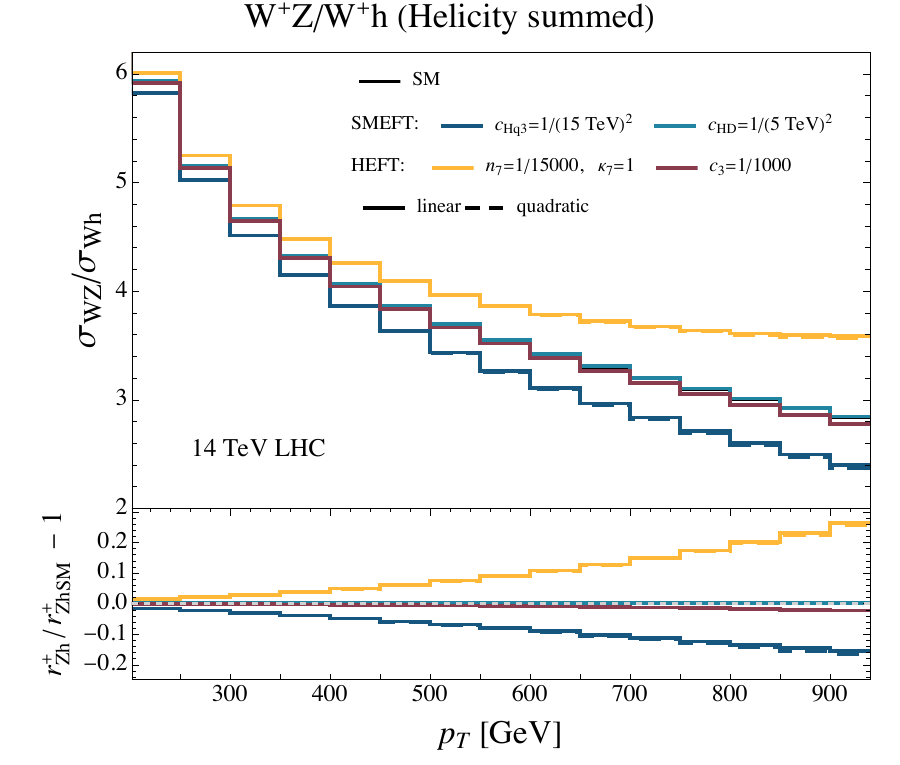}

\caption{Transverse-momentum distributions for (top) $W^+Z$ production, (middle) $W^+h$ production, and (bottom) $W^+Z/W^+h$ ratio. The upper panels show $d\sigma/dp_T^W$ for the $W^+Z$ and $W^+h$ channels  and the bottom row shows the cross section ratio $r^+_{Zh}$ defined in Eq.~(\ref{eq:rHad}), the lower panels show the corresponding deviations relative to the SM. The left column corresponds to fully longitudinal final states, and the right column to helicity-summed final states. In all panels, the SM is shown in black, $C_{Hq}^{(3)}=(15~{\rm TeV})^{-2}$ in dark blue, $C_{HD}=(5~{\rm TeV})^{-2}$ in light blue, $n_7=1/15000$ in yellow, and $c_3=1/1000$ in maroon. 
}
\label{fig:pT}
\end{figure*}

From the discussion in the previous section, it is clear that the ratio of rates between $W^\pm_LZ_L$ and $W^\pm_Lh$ is sensitive to whether EW symmetry is realized linearly (SMEFT) or non-linearly (HEFT). In dimension-6 SMEFT, the ratio of these rates is expected to converge to one,  whereas there is no such expectation in HEFT.  Additionally, in the fully leptonic mode, the joint polarization of $W^\pm_LZ_L$ has already been observed at the LHC~\cite{ATLAS:2022oge,ATLAS:2024qbd}; and recent phenomenological work also indicates that the $W^\pm_L h$ channel is a viable target for a dedicated LHC analysis \cite{Colyer:2025ehv}. 

Due to its theoretical and experimental appeal, in this section, we discuss the ratio of $W^\pm Z$ and $W^\pm h$ at the hadronic cross section level. 
We define the differential ratio: 
\begin{align}
r^\pm_{Zh}
&\equiv
\frac{d\sigma(pp\to W^\pm Z)/dp_T^W}
     {d\sigma(pp\to W^\pm h)/dp_T^W}\,,
\\[2mm]
r_{Zh}
&\equiv
\frac{
\sum_{W^\pm} d\sigma(pp\to W^\pm Z)/dp_T^W
}{
\sum_{W^\pm} d\sigma(pp\to W^\pm h)/dp_T^W
}\,,
\label{eq:rHad}
\end{align}
 where $p_T^W$ is the transverse momentum of the final state $W^\pm$; and the notation $r^+_{Zh}$ means $W^+$-only differential ratio, $r^-_{Zh}$ means $W^-$-only differential ratio, and no superscript $r_{Zh}$ means that the charges are summed in the numerator and denominator separately.
Figure~\ref{fig:pT} shows (top) $d\sigma(pp\rightarrow W^+ Z)/dp_T^W$, (middle) $d\sigma(pp\rightarrow W^+ h)/dp_T^W$, and (bottom) $r^+_{Zh}$ for (left) longitudinally polarized final states and (right) helicity summed final state.
We used the CTEQ6L1 parton distribution functions (PDFs)~\cite{Pumplin:2002vw}, through the LHAPDF library~\cite{Buckley:2014ana} via the \textsc{ManeParse} interface~\cite{Clark:2016jgm} for Mathematica. Our results are validated with independent simulations performed using \textsc{MadGraph5\_aMC@NLO}~\cite{Alwall:2014hca}.  We use the parameter choices as in Eq.~(\ref{eq:params}).

As can be seen, for our chosen parameter points, the quadratic contributions are much less than the linear EFT contributions over the range $p_T^W\leq 1$~TeV for both longitudinally polarized and helicity summed gauge bosons. Additionally, for the individual $p_T^W$ distributions, both SMEFT and HEFT deviate from SM predictions. However, as expected, in the bottom panel we show that the ratio $r^+_{Zh}$ is more sensitive to HEFT operators for the longitudinal case. In the SM and SMEFT, this ratio converges to one, while for HEFT it deviates from SM/SMEFT behavior at high-$p_T^W$, for our parameter points. By contrast, in the helicity-summed distributions, SMEFT and HEFT produce similar deviations from the SM in $r^+_{Zh}$, making it difficult to distinguish the two scenarios without extra polarization information. The corresponding \(W^-\) distributions differ from the \(W^+\) because different proton PDFs enter the charge-conjugate initial states. However, the numerator and denominator
of \(r^\pm_{Zh}\) contain the same charged-current luminosity. As discussed in the previous section, in the SM and dimension-6 SMEFT
the ratio of the longitudinal partonic cross sections approaches one at
high energy, independently of the \(W\) charge.
Therefore, both
\(r^+_{Zh}\) and \(r^-_{Zh}\) approach one in the high-$p_T^W$ limit.
The same conclusion also holds for the charge-summed ratio \(r_{Zh}\). Hence, {\it tagging gauge boson polarizations and comparing the rates for $W^\pm_L Z_L$ and $W^\pm_L h$ at high energies provides a probe of the linear vs. non-linear realizations of the EW symmetry}. 

We evaluate the projected sensitivities of the SMEFT operator $C_{Hq}^{(3)}$ and the HEFT operator $n_7^Q$ at the $\sqrt{S}=14$~TeV LHC with integrated luminosity $\mathcal{L}=3~\mathrm{ab}^{-1}$. We will find sensitivity using charge summed \(W_L^{\pm}Z_L\) only analysis and the ratio \(r_{Zh}\) to illustrate how the ratio observable can separate linear vs non-linear realizations of EW symmetry. We denote the longitudinal fraction as  
\begin{equation}
f_{00}^{WX}
\equiv
\frac{\sum_{W^\pm}d\sigma(pp\to W^\pm_L X_L)/dp_T^W}
     {\sum_{W^\pm}d\sigma(pp\to W^\pm X)/dp_T^W}\, ,
\end{equation}
where $X=Z,h$, and the charge conjugate processes are summed in the numerator and denominator separately. In our analysis, we choose the following $p_T^W$ bins 
\begin{equation}
100~\mathrm{GeV}<p_T^W<200~\mathrm{GeV},
\qquad
p_T^W>200~\mathrm{GeV}.
\end{equation}
Using SM predictions as the reference point, we find
\begin{align}
f_{00}^{WZ(\rm SM)}=\begin{cases}
0.14&~{\rm if}~100~{\rm GeV}<p_T^W<200~{\rm GeV}\\
0.22&~{\rm if}~200~{\rm GeV}<p_T^W
\end{cases}.
\end{align}
To estimate the sensitivity, we start from the current LHC uncertainties for the two bins. The statistical and systematic uncertainties for the measurement of $f_{00}$ in the combined \(W^{\pm}Z\to \ell^{\pm}\nu\,\ell'\ell'\) (\(\ell,\ell'=e,\mu\)) production at \(\sqrt{S}=13~{\rm TeV}\) with an integrated luminosity of
\(140~{\rm fb}^{-1}\) are reported in~\cite{ATLAS:2024qbd}: \footnote{The uncertainties quoted in Ref.~\cite{ATLAS:2024qbd} are obtained with the event selection and fiducial cuts used in the ATLAS analysis. We use them as representative estimates for our projection.}
\begin{align}
&\delta_{\rm stat}^{\rm LHC}=\begin{cases}
0.03&~{\rm if}~100~{\rm GeV}<p_T^W<200~{\rm GeV}\\
0.08&~{\rm if}~200~{\rm GeV}<p_T^W
\end{cases}\\
&\delta_{\rm syst}^{\rm LHC}=\begin{cases}
0.02&~{\rm if}~100~{\rm GeV}<p_T^W<200~{\rm GeV}\\
0.02&~{\rm if}~200~{\rm GeV}<p_T^W
\end{cases}
\end{align}
To estimate the HL-LHC sensitivity, we perform a simple extrapolation of the above ATLAS uncertainties. We assume the benchmark scenario where the systematic uncertainty is reduced by a factor of 1/2 \cite{ATLAS:2025eii}, and rescale the statistical uncertainties according to luminosity and SM rates:
\begin{subequations}
\begin{align}
\delta^{\rm HL}_{\rm syst}&=\frac{1}{2}\delta^{\rm LHC}_{\rm syst}\\
&=\begin{cases}
0.01&~{\rm if}~100~{\rm GeV}<p_T^W<200~{\rm GeV}\\
0.01&~{\rm if}~200~{\rm GeV}<p_T^W
\end{cases},\nonumber\\
\delta^{\rm HL}_{\rm stat}
&=
\delta^{\rm LHC}_{\rm stat}\sqrt{
\frac{\mathcal L_{\rm LHC}\,\sigma^{\rm SM}_{13}}
     {\mathcal L_{\rm HL}\,\sigma^{\rm SM}_{14}}
}\\
&=\begin{cases}
0.006 &~{\rm if}~100~{\rm GeV}<p_T^W<200~{\rm GeV}\\
0.016 &~{\rm if}~200~{\rm GeV}<p_T^W
\end{cases},\nonumber
\end{align}
\end{subequations}
where $\mathcal L_{\rm LHC}=140~\mathrm{fb}^{-1}$, $\mathcal L_{\rm HL}=3000~\mathrm{fb}^{-1}$, and $\sigma^{\rm SM}_{13}$ ($\sigma^{\rm SM}_{14}$) is the SM cross section at the $\sqrt{S}=13$ TeV (14 TeV) LHC.  The rescaling is performed bin-by-bin. 

\begin{figure*}[t]
    \centering
    \includegraphics[width=0.48\textwidth]{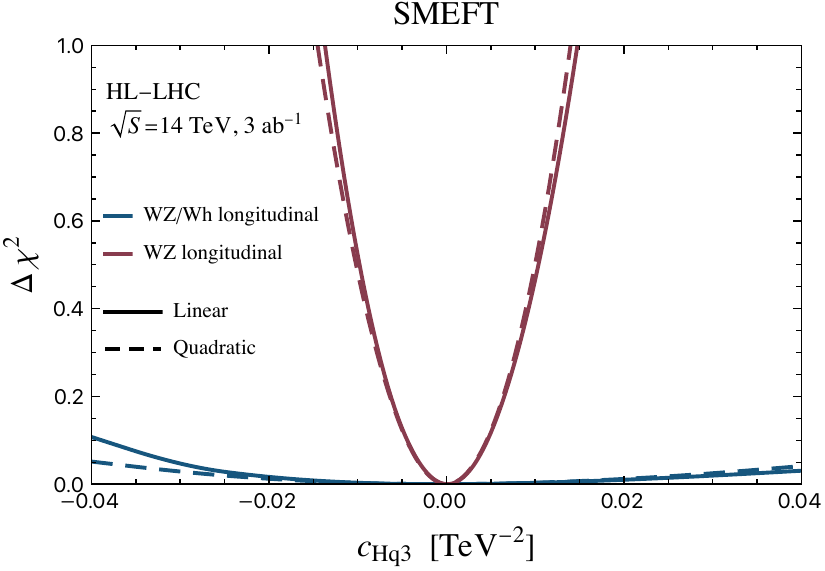}%
    \hfill \includegraphics[width=0.48\textwidth]{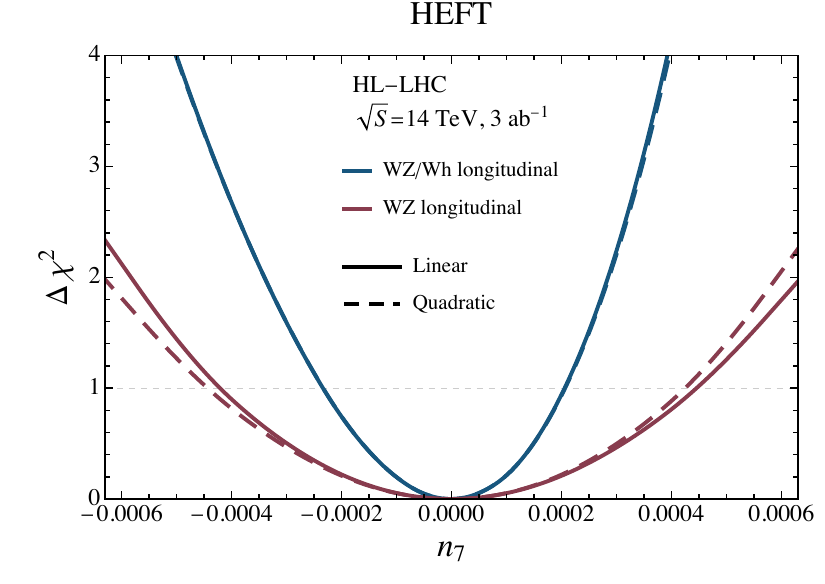}%
    \caption{$\Delta\chi^2$ for a one parameter fit as a function of the SMEFT operator $C_{Hq}^{(3)}$ (left) and the HEFT coefficient $n_7^Q$ with $\kappa_7=1$ (right) at the HL-LHC, obtained from the charge summed $W^\pm_LZ_L$ channel alone (maroon) and from the ratio $r_{Zh}$ (blue) defined in Eq.~(\ref{eq:rHad}). Solid curves include only the linear EFT contribution and dashed curves include quadratic contributions.}
    \label{fig:chi2}
\end{figure*}

Combining statistical and systematic uncertainties in quadrature, we obtain the total projected uncertainties
\begin{align}
\delta^{\rm HL}_{ WZ,i}
&=
\sqrt{
\left(\delta_{\rm stat}^{\rm HL}\right)^2
+
\left(\delta_{\rm syst}^{\rm HL}\right)^2
}
\\
&=\begin{cases}
0.012&~{\rm if}~100~{\rm GeV}<p_T^W<200~{\rm GeV}\\
0.019&~{\rm if}~200~{\rm GeV}<p_T^W
\end{cases}\nonumber
\end{align}

For the combined $W^\pm_L Z_L$ only analysis we compute the $\Delta \chi^2$
\begin{equation}
\Delta\chi^2_{WZ}
=
\sum_{i}
\frac{
\left[
f_{00,i}^{WZ(\rm EFT)}
-
f_{00,i}^{WZ(\rm SM)}
\right]^2
}{
\left(\delta_{WZ,i}^{\rm HL}\right)^2
},
\end{equation}
where we sum over the $p_T^W$ bins, $f_{00,i}^{WZ}$ is the longitudinal fraction in the $i$th $p_T^W$ bin, $\delta^{\rm HL}_{WZ,i}$ is projected uncertainty of the longitudinal fraction measurement in the $i$th bin, and the superscript ${\rm EFT}$ (${\rm SM}$) indicate the prediction in HEFT/SMEFT (SM).

To compare the sensitivity of the charge summed $W_L^{\pm} Z_L/W^{\pm}_L h$ ratio 
with the \(W_L^{\pm} Z_L\) only analysis, we define the binned ratio
\begin{align}
r_{Zh,i}
&=
\frac{\sum_{W^{\pm}}\sigma_i(pp\to W_L^{\pm} Z_L)}
     {\sum_{W^{\pm}}\sigma_i(pp\to W_L^{\pm} h)} \\
&=
\frac{f_{00,i}^{WZ}}{f_{0,i}^{Wh}}\,
\frac{\sum_{W^{\pm}}\sigma_i(pp\to W^{\pm} Z)}
     {\sum_{W^{\pm}}\sigma_i(pp\to W^{\pm} h)},
\end{align}
where $\sigma_i$ denotes the cross section integrated over the $i$th $p_T^W$ bin, and the charge conjugate processes are summed in the numerator and denominator separately. The uncertainty of $r_{Zh,i}$ is estimated as
\begin{equation}
\delta r^{{\rm HL}}_{Zh,i}
=
r_{Zh,i}^{(\rm SM)}
\sqrt{
\left(\epsilon_{{ WZ},i}^{\rm HL}\right)^2
+
\left(\epsilon_{{Wh},i}^{\rm HL}\right)^2
},
\end{equation}
where the total fractional uncertainties are
\begin{subequations}
\begin{eqnarray}
\epsilon_{WZ,i}^{\rm HL}&=&\frac{\delta \sum_{W^{\pm}}\sigma^{\rm HL}_i(pp\rightarrow W^{\pm}_LZ_L)}{\sum_{W^{\pm}}\sigma_i^{({\rm SM)}}(pp\rightarrow W^{\pm}_LZ_L)}\\
\epsilon_{Wh,i}^{\rm HL}&=&\frac{\delta \sum_{W^{\pm}}\sigma^{\rm HL}_i(pp\rightarrow W^{\pm}_Lh)}{\sum_{W^{\pm}}\sigma^{({\rm SM})}_i(pp\rightarrow W^{\pm}_Lh)},
\end{eqnarray}
\end{subequations}
and the superscript ${\rm HL}$ denotes uncertainties at the HL-LHC.

Ref.~\cite{Colyer:2025ehv} projects an inclusive measurement of the longitudinal $W_L^\pm h$ cross section at $3000~\mathrm{fb}^{-1}$ with a precision of approximately $10\%$ in \(W^{\pm}h\to \ell^{\pm}\nu\,\gamma \gamma\) (\(\ell=e,\mu\)) production. Since no dedicated study of the longitudinal fraction is currently available in $p_T^W$ bins, we assign a total fractional uncertainty of $\epsilon_{{ Wh},i}^{\rm HL}=15\%$ in each of the two bins. For the $WZ$ channel, current fully leptonic measurements show a total cross section uncertainty of $\delta \sigma^{\rm HL}(pp\rightarrow W^\pm Z)/\sigma(pp\rightarrow W^\pm Z)\sim 5\%$--$6\%$ in the bins relevant for our analysis~\cite{ATLAS:2025edf}. We project this to the HL-LHC by reducing it by a factor of two. Then we find

\begin{align}
\epsilon_{WZ,i}^{\rm HL}&=\sqrt{\left(\frac{\delta_{WZ,i}^{\rm HL}}{f_{00,i}^{\rm WZ}}\right)^2
+
\left(\frac{\delta \sum_{W^{\pm}}\sigma^{\rm HL}_i(pp\rightarrow W^{\pm} Z)}{\sum_{W^{\pm}}\sigma_i(pp\rightarrow W^{\pm} Z)}\right)^2}\nonumber\\
&=\begin{cases}
0.088&~{\rm if}~100~{\rm GeV}<p_T^W<200~{\rm GeV}\\
0.091 &~{\rm if}~200~{\rm GeV}<p_T^W.
\end{cases}
\end{align}

The $\Delta\chi^2$ test for the ratio observable is defined as
\begin{equation}
\Delta\chi^2_{WZ/Wh}
=
\sum_{i}
\frac{
\left[
r_{Zh,i}^{(\rm EFT)}
-
r_{Zh,i}^{(\rm SM)}
\right]^2
}{
(\delta r^{{\rm HL}}_{Zh,i})^2
},
\end{equation}
where we sum over the same two $p_T^W$ bins as in the $W^{\pm}_LZ_L$ only analysis.

Fig.~\ref{fig:chi2} compares the $\Delta\chi^2$ sensitivities obtained from the joint $W_L^{\pm}Z_L$ only analysis and from the $r_{Zh}$ ratio analysis for the SMEFT operator $C_{Hq}^{(3)}$ and the HEFT operator $n_7^Q$ with $\kappa_7=1$. In both cases, we include the linear interference term and the quadratic contribution. The figure shows that, for both SMEFT and HEFT, the dominant sensitivity comes from the interference term over the parameter range considered. 

The key observation is that the ratio observable is much more sensitive to HEFT operators such as $n_7^Q$, than to the SMEFT operator $C_{Hq}^{(3)}$. In the SMEFT case, the $r_{Zh}$ ratio is \textit{significantly} less sensitive than the charge summed $W_L^{\pm}Z_L$ only analysis. This occurs because the leading high energy effects of $C_{Hq}^{(3)}$ are correlated between the $W_L^{\pm}Z_L$ and $W_L^{\pm}h$ channels at high-$p_T$. For the $n_7^Q$, the ratio observable improves the sensitivity relative to the $W^\pm_LZ_L$ only analysis. This enhancement occurs because $n_7^Q$ and $\kappa_7 n_7^Q$ enter the high energy amplitudes for $W^\pm_LZ_L$ and $W^\pm_Lh$ with opposite signs relative to the SM contributions, as seen in Eqs.~\ref{eq:HEFTWZ} and \ref{eq:HEFTWh}. For $\kappa_7>0$, this leads to constructive interference in one channel and destructive interference in the other. Note that, as mentioned previously, considering $n_7^Q$, this ratio observable goes to one in HEFT only if $\kappa_7$ is tuned to $-1$ due to the interference between HEFT and the SM. SMEFT does not require such a tuning of $C_{Hq}^{(3)}$.

Table~\ref{tab:fit} summarizes the comparison of sensitivity for SMEFT coefficients $C_{Hq}^{(3)}$, $C_{HD}$ and HEFT coefficients $n_7^Q,c_{3}$. The sensitivity to $C_{HD}$ is considerably weaker than that to $C_{Hq}^{(3)}$, since $C_{HD}$ does not generate an $\mathcal{O}(s)$ contribution. However, the 
ratio $r_{Zh}$ improves the bounds on $C_{HD}$ as it introduces custodial-violating effects in the neutral
electroweak sector. As a result, its contributions to
$W^\pm_LZ_L$ and $W^\pm_Lh$ are not correlated, and the ratio is
sensitive to the resulting mismatch between the two channels. Additionally, the bounds on $c_3$ from $W^\pm_LZ$ only and $W^\pm_LZ_L/W^\pm_Lh$ ratio are similar. This is due to $c_3$ having $\mathcal{O}(s)$ behavior in $W^\pm_LZ_L$ production but not $W^\pm_Lh$ production. Hence, the ratio does not introduce any additional behavior that is not already taken into account by measuring $W^\pm_LZ_L$ only. However, in SMEFT the leading
energy growing effects are correlated between $W^\pm_LZ_L$ and
$W^\pm_Lh$ production. Therefore, comparing the rates of the two processes at high energy is still necessary to elucidate the nature of EW symmetry.

\begin{table*}[t]
  \centering
  \large
  \setlength{\tabcolsep}{8pt}
  \renewcommand{\arraystretch}{0.8}
  \begin{tabular}{llcc}
    \toprule
    Observable & EFT treatment & $68\%$ CL & $95\%$ CL \\
    \midrule

    \multicolumn{4}{c}{\textit{SMEFT coefficients}}\\
    \midrule
    \multicolumn{4}{c}{\textbf{$C_{Hq}^{(3)}\;[\mathrm{TeV}^{-2}]$}}\\
    \midrule
    $W^{\pm}Z$ only      & Linear    
      & $(-1.37,\,1.47)\times 10^{-2}$ 
      & $(-2.66,\,3.05)\times 10^{-2}$ \\
    $W^{\pm}Z$ only      & Quadratic 
      & $(-1.46,\,1.40)\times 10^{-2}$ 
      & $(-3.01,\,2.76)\times 10^{-2}$ \\
    $W^{\pm}Z/W^{\pm}h$ ratio  & Linear    
      & $(-1.3,\,4.5)\times 10^{-1}$ 
      & $(-2.3,\,9.3)\times 10^{-1}$ \\
    $W^{\pm}Z/W^{\pm}h$ ratio  & Quadratic 
      & $(-2.7,\,3.2)\times 10^{-1}$ 
      & $(-5.5,\,6.5)\times 10^{-1}$ \\
    \midrule

    \multicolumn{4}{c}{\textbf{$C_{HD}\;[\mathrm{TeV}^{-2}]$}}\\
    \midrule
    $W^{\pm}Z$ only      & Linear    
      & $(-9.18,\,10.41)$
      & $(-18.90,\,20.66)$ \\
    $W^{\pm}Z$ only      & Quadratic 
      & $(-9.18,\,11.89)$
      & $(-17.36,\,23.80)$ \\
    $W^{\pm}Z/W^{\pm}h$ ratio  & Linear    
      & $(-2.16,\,1.98)$
      & $(-4.34,\,3.84)$ \\
    $W^{\pm}Z/W^{\pm}h$ ratio  & Quadratic 
      & $(-2.16,\,1.97)$
      & $(-4.34,\,3.86)$ \\
    \midrule

    \multicolumn{4}{c}{\textit{HEFT coefficients}}\\
    \midrule
    \multicolumn{4}{c}{\textbf{$n_7^Q \, (\kappa_7=1)$}}\\
    \midrule
    $W^{\pm}Z$ only      & Linear    
      & $(-4.19,\,4.45)\times 10^{-4}$  
      & $(-8.18,\,9.12)\times 10^{-4}$ \\
    $W^{\pm}Z$ only      & Quadratic 
      & $(-4.43,\,4.23)\times 10^{-4}$ 
      & $(-9.09,\,8.36)\times 10^{-4}$ \\
    $W^{\pm}Z/W^{\pm}h$ ratio  & Linear    
      & $(-3.24,\,2.76)\times 10^{-4}$  
      & $(-7.11,\,5.07)\times 10^{-4}$ \\
    $W^{\pm}Z/W^{\pm}h$ ratio  & Quadratic 
      & $(-3.24,\,2.77)\times 10^{-4}$ 
      & $(-7.17,\,5.15)\times 10^{-4}$ \\
    \midrule

    \multicolumn{4}{c}{\textbf{$c_3$}}\\
    \midrule
    $W^{\pm}Z$ only      & Linear    
      & $(-1.6,\,1.6)\times 10^{-2}$
      & $(-3.4,\,3.1)\times 10^{-2}$ \\
    $W^{\pm}Z$ only      & Quadratic 
      & $(-1.6,\,1.6)\times 10^{-2}$
      & $(-3.2,\,3.5)\times 10^{-2}$ \\
    $W^{\pm}Z/W^{\pm}h$ ratio  & Linear    
      & $(-2.4,\,2.4)\times 10^{-2}$
      & $(-5.0,\,5.0)\times 10^{-2}$ \\
    $W^{\pm}Z/W^{\pm}h$ ratio  & Quadratic 
      & $(-2.3,\,2.7)\times 10^{-2}$
      & $(-4.3,\,6.4)\times 10^{-2}$ \\
    \bottomrule
  \end{tabular}
  \caption{\label{tab:fit}
  Single-parameter bounds from the $\Delta\chi^2$ analysis for selected SMEFT and HEFT coefficients, using the charge summed $W^{\pm}Z$ only analysis and the $W^{\pm}Z/W^{\pm}h$ ratio analysis. The SMEFT coefficients are shown in units of $\mathrm{TeV}^{-2}$, and HEFT coefficients are dimensionless following the normalization in the text. Bounds are shown for both linear and quadratic EFT treatments of the varied coefficient.
  For the $n_7^Q$ fit we set $\kappa_7=1$.
  }
\end{table*}

\section{Conclusions}
\label{sec:conc}

While many studies comparing SMEFT and HEFT focus on comparing single and multi-Higgs production or longitudinal gauge boson scattering~\cite{Chang:2019vez,Abu-Ajamieh:2020yqi,Gomez-Ambrosio:2022qsi,Domenech:2022uud,Englert:2023uug,Delgado:2023ynh,Remmen:2024hry,Domenech:2025gmn} to distinguish the two theories, our study shows that comparing zero and one Higgs production via fermion-antifermion annihilation can also distinguish whether EW symmetry is linearly or non-linearly realized. Whether or not the Higgs boson lives in a doublet or is a singlet is reflected in high energy behavior of ratios of amplitudes for $f\bar{f}'\rightarrow V_LV'_L$ and $f\bar{f}'\rightarrow V_Lh$. As we showed explicitly, both the SM and dimension-6 SMEFT predict certain ratios for longitudinal di-boson amplitudes to approach one in the high energy limit, while that is not necessarily true in HEFT. {\it In both the SM and SMEFT, this can be understood via the Goldstone boson equivalence theorem together with EW restoration}. In the EW restored SM, the pair production of Goldstone bosons, $GG'$, or a Goldstone boson plus Higgs, $Gh$, via fermion-antifermion annihilation proceeds through the $s$-channel exchange of hypercharge or $SU(2)_L$ gauge bosons. That is, the high energy $GG'$ and $Gh$ pair production in the SM proceeds through the product of quark and Higgs currents.  

In SMEFT, the quadratic energy growth in Goldstone and Higgs boson pair production proceeds through the operators in Eqs.~(\ref{eq:QHq}-\ref{eq:QHud}), which are just the same products of quark and Higgs field currents as in the SM. As we showed, these products can be used to derive relationships among Goldstone boson pair production that predict the SM and SMEFT relationships in Table~\ref{tab:AmpRatEWR} via the Goldstone boson equivalence theorem. In HEFT, there still are expected approximate relationships among $f\bar{f}'\rightarrow V_LV'_L$ amplitudes, up to energy growing custodial symmetry violations. In general, however, there are no relationships between $f\bar{f}'\rightarrow V_LV'_L$ and $f\bar{f}'\rightarrow V_L h$ amplitudes in HEFT. Hence, the ratios of $V_LV_L'$ and $V_Lh$ amplitudes provide a test of linear vs. non-linear EW symmetry.

For a complete test, all the high energy di-boson correlations predicted by a linearly realized electroweak symmetry should be measured. While there are ratios of amplitudes for all di-boson final states that approach one at high energy in the SM and SMEFT, these depend on initial state fermion helicity. Once fermion helicities are summed over, the ratio of $W^\pm_L Z_L$ and $W^\pm_L h$ rate still approaches one, while ratios of other di-boson rates can deviate from the SM in both SMEFT and HEFT.  This is shown in Figs.~\ref{fig:xsectRat} and~\ref{fig:pT}, at both the partonic and hadronic level. Additionally, in the top two plots of Fig.~\ref{fig:xsectRat} and the bottom two plots of Fig.~\ref{fig:pT}, we showed that polarization tagging of final state gauge bosons is important. When the gauge boson helicities are summed, both SMEFT and HEFT can cause similar deviations from the SM predictions of the ratio of $W^\pm Z$ and $W^\pm h$ production. Hence, to test the linear vs. non-linear realization of EW symmetry, tagging longitudinally polarized gauge bosons is crucial for unambiguous predictions in dimension-6 SMEFT.

In addition to the theoretical prediction, the longitudinal gauge boson production $W^\pm_LZ_L$ has already been observed at the LHC at $7.1\sigma$~\cite{ATLAS:2022oge,ATLAS:2024qbd} and there are ongoing measurements of $W^\pm h$~\cite{ATLAS:2024yzu,CMS:2026fsx,CMS:2023vzh}. Hence, with the addition of polarization tagging in $Wh$ production, current measurements could be interpreted as a test of linear vs. non-linear realizations of the EW symmetry. Indeed, using current measurements and projections of $W^\pm_Lh$ measurements~\cite{Colyer:2025ehv}, we projected bounds on Wilson coefficients of SMEFT and HEFT using measurements of $W^\pm_LZ_L$ by themselves and the ratio of $W^\pm_LZ_L$ and $W^\pm_L h$. This showed unambiguously that comparing these rates can provide stronger constraints on some HEFT parameters since the ratios do not generically approach one. Even when the ratio measurement does not improve a bound on a HEFT coefficient, the comparison of rates is still necessary to test linear vs. non-linear realizations of EW symmetry since dimension-6 SMEFT has predictions for how different rates correlate. Further exploration on extracting gauge boson polarization information, such as the polarization density matrices, using both inclusive and exclusive observables, would enable deeper tests of the SM, EWSB, and EW restoration.


\begin{acknowledgments}
We thank Tao Han, Ian Low, Adam Martin, Patrick Meade,  and Jure Zupan for helpful discussions. 
IML is supported in part by the United States Department of Energy grant number DE-SC001798.  Z.L. and I.M. are supported by the Department of Energy under Grant No.~DE-SC0011842 at the University of Minnesota. Z.L. is supported in part by a Sloan Research Fellowship from the Alfred P. Sloan Foundation at the University of Minnesota.
\end{acknowledgments}
 

\bibliographystyle{apsrev4-2}
\bibliography{ref}
\appendix

\end{document}